%% file: ANA-EXOT-2020-26-PAPER.tex
\documentclass[PAPER, cernpreprint, atlasdraft=false, USenglish, texmf, orcidlogo]{atlasdoc}
\usepackage{atlaspackage}
\usepackage{atlasbiblatex}

\usepackage[bsm=true,jetetmiss=true]{atlasphysics}
\graphicspath{{logos/}{figures/}}

\usepackage{ANA-EXOT-2020-26-PAPER-defs}
\usepackage{multicol}
\usepackage{multirow}
\usepackage{rotating}
\usepackage{tabularx}
\newcolumntype{Y}{>{\centering\arraybackslash}X}
\AtlasTitle{Search for dark matter produced in association with a dark Higgs boson in the $b\bar{b}$ final state using $pp$ collisions at $\sqrt{s}=13$~TeV with the ATLAS detector}

\AtlasAbstract{%
A search is performed for dark matter particles produced in association with a resonantly produced pair of $b$-quarks with $30< m_{bb}<150$~GeV using 140~fb$^{-1}$ of proton--proton collisions at a center-of-mass energy of 13~TeV recorded by the ATLAS detector at the LHC. This signature is expected in extensions of the Standard Model predicting the production of dark matter particles, in particular those containing a dark Higgs boson $s$ that decays into $b\bar{b}$.
The highly boosted $s \rightarrow b\bar{b}$ topology is reconstructed using jet reclustering and a new identification algorithm.
This search places stringent constraints across regions of the dark Higgs model parameter space that satisfy the observed relic density, excluding dark Higgs bosons with masses between $30$ and $150$~GeV in benchmark scenarios with $Z'$ mediator masses up to $4.8$~TeV at 95\% confidence level.

}

\AtlasRefCode{EXOT-2020-26}
\PreprintIdNumber{CERN-EP-2024-170}

\arXivId{2407.10549}

\AtlasJournal{Phys.\ Rev.\ Lett.}
\AtlasJournalRef{Phys.\ Rev.\ Lett.\ 134 (2025) 121801}
\AtlasDOI{https://doi.org/10.1103/PhysRevLett.134.121801}

\hypersetup{pdftitle={ATLAS document},pdfauthor={The ATLAS Collaboration}}

\begin{document}

\maketitle

Multiple astrophysical observations~\cite{Bertone:2010zza,Feng:2010gw,Porter:2011nv,Bertone:2017adx} indicate that a large fraction of the matter density of the universe is in the form of dark matter (DM). Its nature is a major open question in physics, for the Standard Model (SM) of particle physics does not provide any suitable DM candidates. Many extensions of the SM propose DM candidates that are stable, neutral, massive weakly interacting particles~\cite{Bertone:2017adx}, which determine the DM relic abundance via thermal freeze-out.
The search for DM candidates is being pursued actively in direct and indirect detection experiments in addition to collider experiments~\cite{PhysRevLett.121.111302,PhysRevLett.117.121303,PhysRevLett.121.081307,PhysRevLett.123.241803,PhysRevD.100.022001,Ackermann_2015,PerezdelosHeros:2020qyt}. Once produced in colliders, DM would be undetected and must be inferred from the imbalance of the transverse momentum \metvec,\footnote{\AtlasCoordFootnote} with magnitude \MET, observed from the detected SM particles.

This Letter presents a novel DM search using a $X+\met$ signature in which $X$ is a hypothetical particle that decays into a $b$-quark pair, $b\bar{b}$. This signature of large \met and resonant $b\bar{b}$ production has not been probed directly for invariant masses $m_{bb}<150~\GeV$, except for when $X$ is the SM Higgs boson, $h$~\cite{EXOT-2018-46,CMS-EXO-16-050}.
Signal regions~(SRs) are defined by requiring significant \MET\ consistent with the presence of DM, in association with a $b\bar{b}$ decay, which is usually the dominant branching fraction for a low mass $X$ with SM Higgs boson couplings. %
The background is dominated by vector-boson production in association with jets, referred to as \vjets, with top quark pair (\ttbar) production also significant at lower \met\ values.
To constrain and improve the modeling of these background contributions, control regions (CRs) are defined that require either a single muon ($\mu$) or a pair of charged leptons $\ell^\pm \ell^\mp$ ($\ell = e, \mu$) in the final state.

The optimization and interpretation of the search is based on a dark Higgs model~\cite{Duerr:2017uap} that explains mass generation for DM particles ($\chi$) through a Higgs mechanism in the dark sector and Yukawa interactions with a new massive dark Higgs boson ($s$).
This model satisfies the observed DM relic density as, when the dark Higgs boson $s$ is lighter than the DM particle $\chi$, additional annihilation channels such as $\chi \chi \rightarrow ss$ can be dominant. Thus it offers a widespread, generic ability to reproduce the observed relic density. %
In this two-mediator DM model, Majorana DM particles interact with the SM via the exchange of new spin-1 ``mediator'' particles carrying a new \Up gauge symmetry (e.g. a new $\Zprime$ gauge boson), %
which can be probed at colliders~\cite{Duerr:2017uap} through $s$-channel processes.
Since large dark sector couplings usually reproduce the relic density, the probability for a $\Zprime$ to radiate a dark Higgs boson can be large.
Annihilation signals in these models are suppressed, as is direct detection sensitivity for  Majorana DM particles; thus colliders provide unique discovery potential.
The key model parameters are the Majorana DM candidate's mass $m_\chi$, the \Zp mass \mZp, the dark Higgs boson mass $m_s$, the two couplings of the \Zp boson to quarks \gq and to DM \gchi, and lastly the mixing angle between the SM and dark Higgs bosons $\theta$.
In this model, the \Zp decays dominantly into DM, which recoils against the dark Higgs boson and its visible decay products.
If the dark Higgs boson is the lightest dark sector state, exploring the dominant decays of low-mass $s$ bosons is vital.
For $\ms < 150$~\GeV, decays into a $b$-quark pair dominate, with the Lorentz boost and collimated decay generating a \merged topology signature of a single large-radius (\largeR) jet containing two $b$-quarks. %
The experimental challenges of this final state are the identification of the massive jet and its $b$-quarks, and maintaining sensitivity to $m_s < 50$~\GeV.
For $\mZp < 2$~\TeV and $\ms > 70$~\GeV, the greater separation of the $b$-quark pair motivates a \resolved topology of two small-radius (\smallR) $b$-quark jets.

The analysis is performed using \Lumi~\ifb\ of \pp collisions at $\sqrt{s}=13~\TeV$ recorded with the ATLAS detector~\cite{PERF-2007-01,ATLAS-TDR-2010-19} in 2015--2018 during good operating conditions~\cite{DAPR-2018-01}. %
Monte Carlo (MC) simulations are used to model the kinematics of SM background processes and the \sDM signal. %
A detailed simulation of the ATLAS detector~\cite{SOFT-2010-01} based on \GEANT~\cite{Agostinelli:2002hh} was used to simulate the detector response for MC event samples. %
Further description of the ATLAS detector and the triggers used, along with
a signal diagram and details of the event simulation configurations used for signal and background processes, can be found in the Appendix. %
Signal simulations for the $pp \rightarrow \Zprime \rightarrow s\chi\chi \rightarrow b\bar{b}\chi\chi$ process in three interpretation scenarios are used to
investigate interesting phase spaces of the model. %
Scenario 1 simulations were generated in the (\mZp,~\ms) plane, covering 30--150~\GeV in \ms and \mZp up to 4~\TeV. Other parameter values were $\mDM=200~\GeV$ to avoid $s\to\chi\chi$ decays, $\gq = 0.25$~\cite{Abercrombie:2015wmb,Abdallah:2015ter}, $\gchi=1.0$ and $\sin \theta = 0.01$~\cite{Duerr:2017uap}, defining a conventional benchmark used in previous studies in different final states~\cite{EXOT-2020-04,EXOT-2018-40,CMS-EXO-21-012}.
Two other scenarios are developed, where the coupling parameter $\gchi$ is varied (instead of being set to unity) to ensure all signal points are compatible with the observed relic density, $\Omega h^2 = 0.12$~\cite{Planck:2018vyg}, calculated using \textsc{MadDM}~\cite{MadDM}.
This requirement usually indicates large $g_\chi$ couplings, especially for larger $\mZp$, which enhances the sensitivity of dark Higgs boson signatures and reduces that of others, e.g. where the $\Zp$ decays into quarks.
Signal simulations for scenario 2 were generated in the (\mZp, \ms) plane, but with $\mDM=900~\GeV$ to enable a match with the relic density across the investigated parameter plane.
Finally, scenario~3 explores the (\mZp, \mDM) plane for a fixed dark Higgs boson mass $\ms=70~\GeV$ that coincides with the highest analysis sensitivity.
The other parameters (\gq and $\sin \theta$) in these scenarios match those of scenario~1.

At least one \pp collision vertex reconstructed from at least two inner detector (ID) tracks with $\pt > 0.5~\GeV$ is required in each event. The vertex with the highest $\sum(\pt)^2$ is designated the primary vertex (PV)~\cite{ATL-PHYS-PUB-2015-026}.
Electrons are reconstructed by matching a cluster of energy in the calorimeter to an ID track. Electron candidates are identified using a likelihood-based method and must satisfy the ``loose'' requirement~\cite{EGAM-2018-01} and have $|\eta|<2.47$.
Muons are reconstructed by matching a track or track segment found in the muon spectrometer to an ID track. Muons must satisfy ``loose'' requirements~\cite{MUON-2018-03} and %
have $|\eta|<2.5$.
Electrons and muons must be isolated
according to the track proximity criteria defined in Ref.~\cite{EXOT-2015-23}.
Hadronic $\tau$-lepton decays are identified by an algorithm based on a boosted decision tree~\cite{ATL-PHYS-PUB-2015-045} that combines
calorimeter and ID information. Events with $\tau$-leptons satisfying $\pt>20~\GeV$ and ``very loose'' requirements~\cite{ATL-PHYS-PUB-2019-033} within $|\eta|=2.5$ %
are rejected. %

\SmallR jets are formed with the \antikt algorithm~\cite{Fastjet,Cacciari:2008gp}, using a radius parameter $R=0.4$, from ID tracks associated with the PV and three-dimensional clusters of calorimeter cells selected by a particle-flow reconstruction algorithm~\cite{PERF-2015-09}. ``Central'' \smallR jets satisfy $|\eta|<2.5$ and $\pt>20~\GeV$ while ``forward'' jets satisfy $2.5<|\eta|<4.5$ and $\pt>30~\GeV$. Corrections for pileup~\cite{Cacciari:2008gn} and the jet energy scale (JES) and resolution (JER)~\cite{JETM-2018-05} are applied. The PV origin of central \smallR jets with $20<\pt<60~\GeV$ and $|\eta|<2.4$ is required, using an associated-track-based discriminant~\cite{PERF-2014-03}. Small-$R$ jets closer than $\DeltaR = 0.2$ to an $e$ or $\mu$ %
are rejected.
Two types of \largeR jets are reconstructed using the \antikt algorithm with radius $R=1.0$. ``Reclustered'' \largeR jets ($J$) are derived by clustering \smallR jets ($j$)~\cite{ATLAS-CONF-2017-062}; these are used for the merged analysis with the intention of capturing the dark Higgs boson decay in full, providing sensitivity and good mass resolution across the full jet mass range (down to $m_J = 30$~\GeV). %
Their flavor content is evaluated through their associated variable-radius (VR) track-jets~\cite{ATL-PHYS-PUB-2017-010,ATL-PHYS-PUB-2017-013} or the \Xbb discriminant score of the corresponding calorimeter \largeR jet, as described below.
``Calorimeter'' large-$R$ jets are clustered from topological clusters calibrated to the hadronic scale using the local hadronic cell weighting (LCW) scheme~\cite{PERF-2014-07}.
All \largeR jets are trimmed~\cite{Krohn:2009th} %
to minimise the impact of pileup and underlying event. The JES and jet mass scale (JMS) of trimmed jets are calibrated following techniques described in Ref.~\cite{JETM-2018-02}.

To suppress contributions from processes that involve light quarks or gluons, two multivariate algorithms are used to identify jets containing $b$-hadrons ($b$-tagging)~\cite{FTAG-2019-07}. %
The algorithm \texttt{DL1r} is used at an operating point evaluated to be 77\% efficient at $b$-jet identification on \ttbar simulation~\cite{FTAG-2018-01}, with a light jet rejection factor of around $200$.
For $m_J<50$~\GeV, this algorithm and operating point is applied to VR track-jets with $\pt>10~\GeV$ and $|\eta|<2.5$ formed from ID tracks using the \antikt algorithm and a \pt-dependent radius parameter. %
It is also applied to the \smallR jets in the resolved channel.
A second, new tagging algorithm \Xbb~\cite{ATL-PHYS-PUB-2020-019,ATL-PHYS-PUB-2021-035}, developed specifically for the $X\to \bbbar$ topology, combines the flavor information of up to three VR track-jets within the large-$R$ jet.
This mass-agnostic neural network %
exploits the powerful tagging capability of individual track-jets and their discriminant correlations, together with the knowledge of the large-$R$ jet kinematics. The algorithm is trained on calorimeter \largeR jets with masses above 50~\GeV, where the axes of the \largeR jet and the reclustered \largeR jet lie within $\DeltaR=1.0$. An operating point evaluated to be 50\% efficient at selecting Higgs bosons with $\pT>250$~\GeV, and a multijet rejection factor of around $110$ is employed.
It is calibrated using $Z(\rightarrow b\bar{b})$+jets and $Z(\rightarrow b\bar{b})$+$\gamma$ data samples in four $\pT$ regions, supported by $\ttbar$ and $g\rightarrow b\bar{b}$ topologies.
It is estimated that the \Xbb algorithm improves the sensitivity by a factor of up to 50\% in expected median discovery significance compared with an \monohbb\ analysis~\cite{EXOT-2018-46} using VR track-jet $b$-tagging, neglecting systematic uncertainties.

The \metvec is computed as the negative vector sum of the transverse momenta of the identified and calibrated physics objects in the event, plus a term accounting for low-energy charged particles, using the ``tight'' operating point defined in Ref.~\cite{JETM-2020-03}.
An object-based \met significance $\mathcal{S}$~\cite{JETM-2020-03} discriminates events with genuine \met produced by neutrinos or possible weakly interacting exotic particles, from those events in which \met is caused by mismeasurements or resolution effects.

The signal is characterized by high \met from the DM particle production, and substantial hadronic activity from $s\to b\bar b$ decays that results in an invariant mass consistent with \ms.
Thus events in the SR are required to have $\met>150~\GeV$, either two $b$-tagged \smallR jets or a \largeR jet containing two $b$-quarks, and no isolated $e$ or $\mu$. %
Events in the SR are rejected if a ``loose'' electron or muon with $\pt>7~\GeV$ is present.

The smallest azimuthal angle between the \metvec and any of the three highest-\pt (leading) \smallR jets is required to be at least $20^\circ$ to reduce the multijet background arising from mismeasured jet momenta.
For signal events the \met and the \pt of the reconstructed dark Higgs boson candidate ($\pt^{jj}$ in the resolved region or $\pt^J$ in the merged region) are correlated through the production process. Their ratio is required to be between $0.8$ and $1.3$ to reduce the contributions from \ttbar and \wjets events.

In the \merged channel, to ensure the decay products are contained in the \largeR\ jet, a $2m_J/\pt^J<0.6$ requirement is applied. The dark Higgs boson candidate jet is required to have at least two non-overlapping VR track-jets associated with it. Events with an additional $b$-tagged VR track-jet not associated with the \largeR jet are rejected to suppress top quark pair production.
In the resolved channel, the dominant background process is \ttbar production. This background is reduced by the variables $\mtb{\min/\max} = \sqrt{2 \ptb{\min/\max} \met (1-\cos \Delta \phi(\ptbvec{\min/\max}, \metvec))}$, and a requirement of $\mtb{\min}>170$~\GeV and $\mtb{\max}>200$~\GeV, where $\ptb{\min}$ and $\ptb{\max}$ are defined as the $\pt$ of the $b$-jet that is closer to (min) or further from (max) \metvec in $\phi$. To suppress \ttbar processes further, the central \smallR jet multiplicity is required to be $\leq 4$. An $\mathcal{S}>12$ requirement is also applied, and results in negligible multijet background.

The largest SR background contributions come from SM \znunujets processes ($48$\%--$60$\%), increasing in higher \met categories. In the merged topology, SM diboson production ($17$\%) and $W(\rightarrow \ell\nu)\!+\!{\mathrm{jets}}$ ($9$\%--$13$\%) provide sub-leading contributions; top quark pair production ($10$\%--$30$\%) and $W(\rightarrow \ell\nu)\!+\!{\mathrm{jets}}$ processes ($13$\%--$15$\%) contribute in the resolved toplogy.   %
Two CRs are defined to improve the modeling of the \vjets background: the single-muon CR (\CRone) enriched in \wjets and \ttbar, and the two-lepton CR (\CRtwo) dominated by \zjets.
The \CRone follows the same selection and \met trigger as the SR, except that events must contain exactly one ``medium'' muon~\cite{PERF-2015-10} with $\pt>27~\GeV$ and no ``loose'' electrons with $\pt>7~\GeV$. It is split into two regions depending on the muon charge to provide additional discrimination between these two backgrounds, due to the larger cross-section for $W^+$ boson production in $pp$ collisions. Events in the \CRtwo are selected using the same requirements as in the SR, except that events must contain exactly two oppositely charged ``loose'' electrons or ``medium'' muons, and satisfy $\mathcal{S}>12$, with an additional requirement that this significance be lower than $5$ when considering \met calculated with the two visible leptons. The leading electon (muon) must fulfill $\pt>27~(25)~\GeV$, while the subleading lepton must satisfy $\pt>7~\GeV$. The dilepton system mass and $\pt$ must be consistent with the $Z$ boson hypothesis of $|m_{\ell\ell} - m_Z|< 10$~\GeV and $p_\text{T}^{\ell\ell} > 150$~\GeV.
To maintain sensitivity to signals generating higher \met values and constrain background processes more effectively, events are further categorized in $\met/\GeV$: $[150,200)$, $[200, 350)$ and $[350,500)$ in the \resolved category and $\met/\GeV$: $[500,750)$, $\ge 750$ in the \merged category. The CRs in the resolved category are divided in the same way. The boundary between \resolved and \merged categories at $500$~\GeV is optimized for search sensitivity. To match the \met kinematics of \vjets processes in the SR,
$\ensuremath{\vec{E}_{\mathrm{T,\mu}}^{\mathrm{miss}}}=\metvec+\vec{p}_\text{T}^\mu$
is used in the \CRone, incorporating the \pt of the $W$ boson. Similarly, the addition of $\vec{p}_\text{T}^{\ell\ell}$ in the \CRtwo provides an analog to the \met in the SR.

Experimental systematic uncertainties affect the reconstruction of the dark Higgs boson candidate. These include uncertainties in the JMS~\cite{JETM-2018-02} and the JES and JER~\cite{JETM-2018-05} of both the \smallR and \largeR jets, and uncertainties in the calibrations of the $b$-jet identification algorithms~\cite{FTAG-2018-01,FTAG-2020-08,FTAG-2019-02}. Uncertainties in the lepton identification efficiencies~\cite{EGAM-2018-01,MUON-2018-03}, \met trigger efficiency, energy scale and resolution~\cite{JETM-2020-03} are found to be negligible, as is the uncertainty in the luminosity~\cite{DAPR-2021-01}.
Theoretical systematic uncertainties originate from the modeling of the signal and major background processes and are %
detailed in the Appendix.

\begin{figure}[!tb]
\centering
\includegraphics[width=0.72\textwidth]{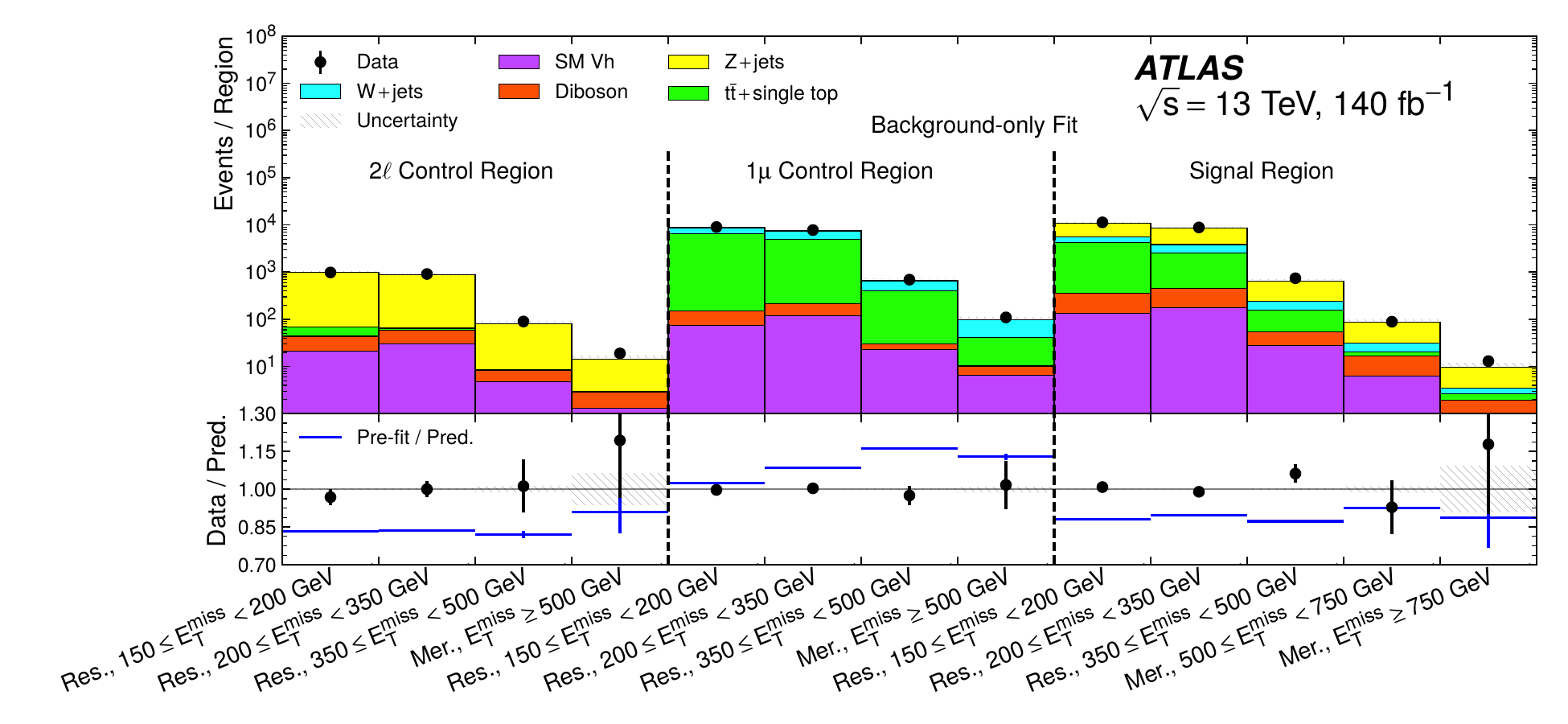}  %
\caption{
Data and predicted SM background yields after a simultaneous background-only fit to each resolved (Res.) and merged (Mer.) SR and CR \met category.
The ratio of the data to the SM expectation is shown in the lower panel; the lines give the ratios of the pre-fit to the post-fit background predictions and the shaded areas indicate the total uncertainty in the predictions.
}
\label{fig:yields-body}
\end{figure}

Limits on DM signals are extracted via a simultaneous maximum-likelihood fit~\cite{Moneta:2010pm,RooFit} of signal and background simulations to the binned candidate mass distributions in the SR and to the event yields in the CRs.
The normalizations of \zjets, \ttbar and \wjets processes are free parameters in the fit and are constrained by the total event yields, considering each \met region separately in the SR and CRs.
Systematic uncertainties are parameterized as nuisance parameters with Gaussian prior probabilities, and constrain the fit templates and normalizations~\cite{HIGG-2013-23}.
Statistical uncertainties in the data are the largest source of uncertainty ($75$\%--$85$\% of the total), with systematic uncertainties, specifically those in the calibration of the \largeR jet $b$-tagging algorithm, becoming more important at larger $\pT$ and \mZp\ values ($20$\%--$47$\%). The largest theoretical uncertainties are in the modeling of \zjets processes ($15$\%--$25$\%), and those associated with the normalization of \vjets processes ($10$\%--$15$\%). At $\ms \approx 50$~\GeV and high $\mZp$, the predominance of \zjets increases the impact of its normalization uncertainty (up to $41$\% of total).
The observed and fitted yields in the SR and CR categories obtained after a simultaneous fit under the hypothesis that only SM contributions are present (``background-only fit'') are shown in Figure~\ref{fig:yields-body}. %
The overall yields in the CRs and the SR are found to be well described by SM expectations, with the fit favoring a mild increase ($\sim 20$\%) of the \zjets contribution.
The pre-fit uncertainties cover the differences between the data and pre-fit background predictions.
Figure~\ref{fig:ms-body} shows the mass distributions $m_{bb}$ of the \s candidate mass in the SR categories after the background-only fit. The MC simulations agree well with the data in the CRs, indicating that \vjets background processes are well modeled.
The data exceed the SM expectation slightly around $\ms = 60$~\GeV in the  $350 \le \met < 500$~\GeV category. Its local significance is 1.6 standard deviations ($\sigma$).
A smaller, very localized excess is also seen near $m_s=130$~\GeV in the same category, and is narrower than the resolution in \ms.
The observed results in the SR indicate that the data are in agreement with SM predictions with no significant evidence of a DM signal.

\begin{figure}[!htb]
\begin{center}
\includegraphics[width=0.32\linewidth]{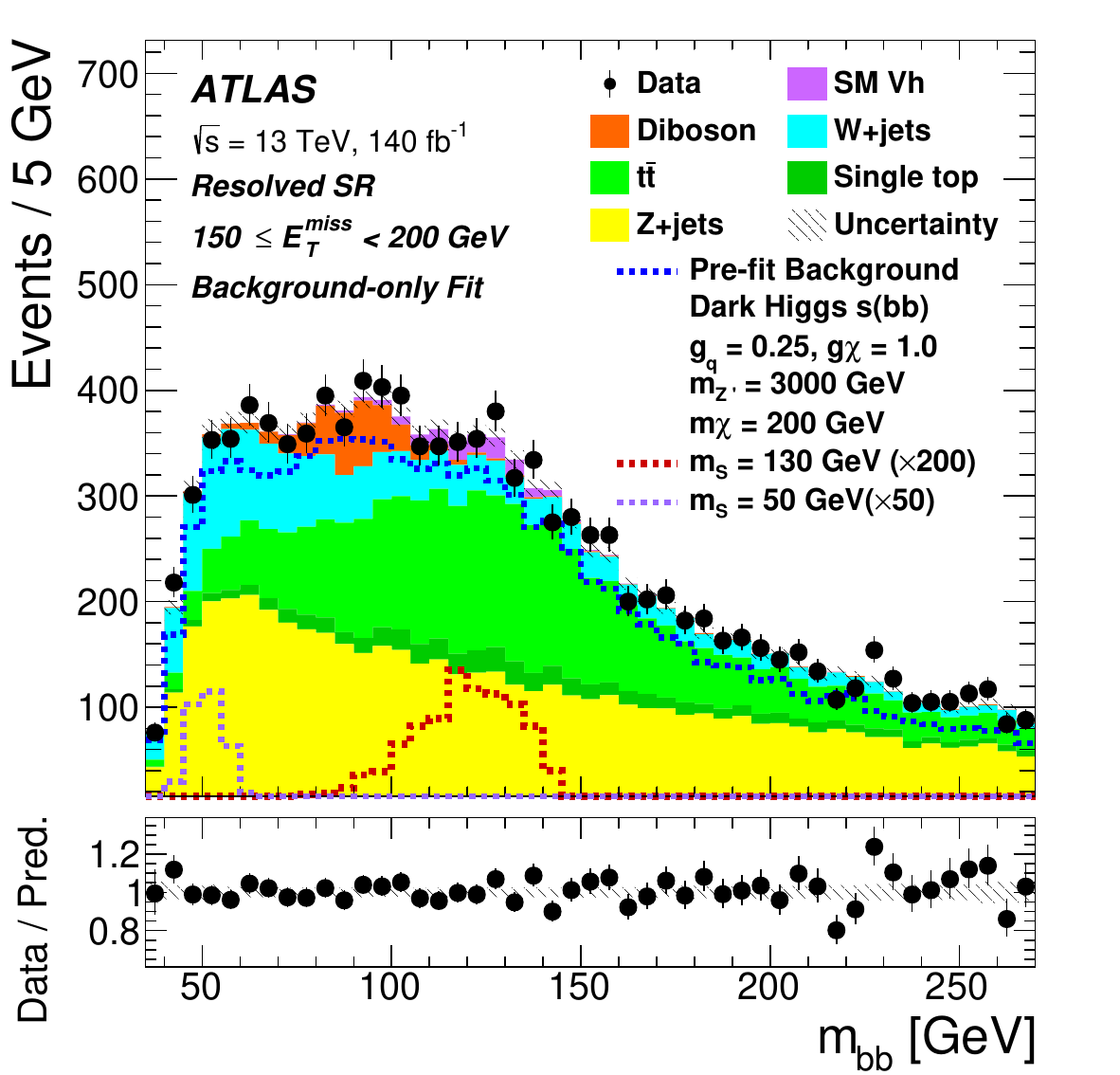} %
\includegraphics[width=0.32\linewidth]{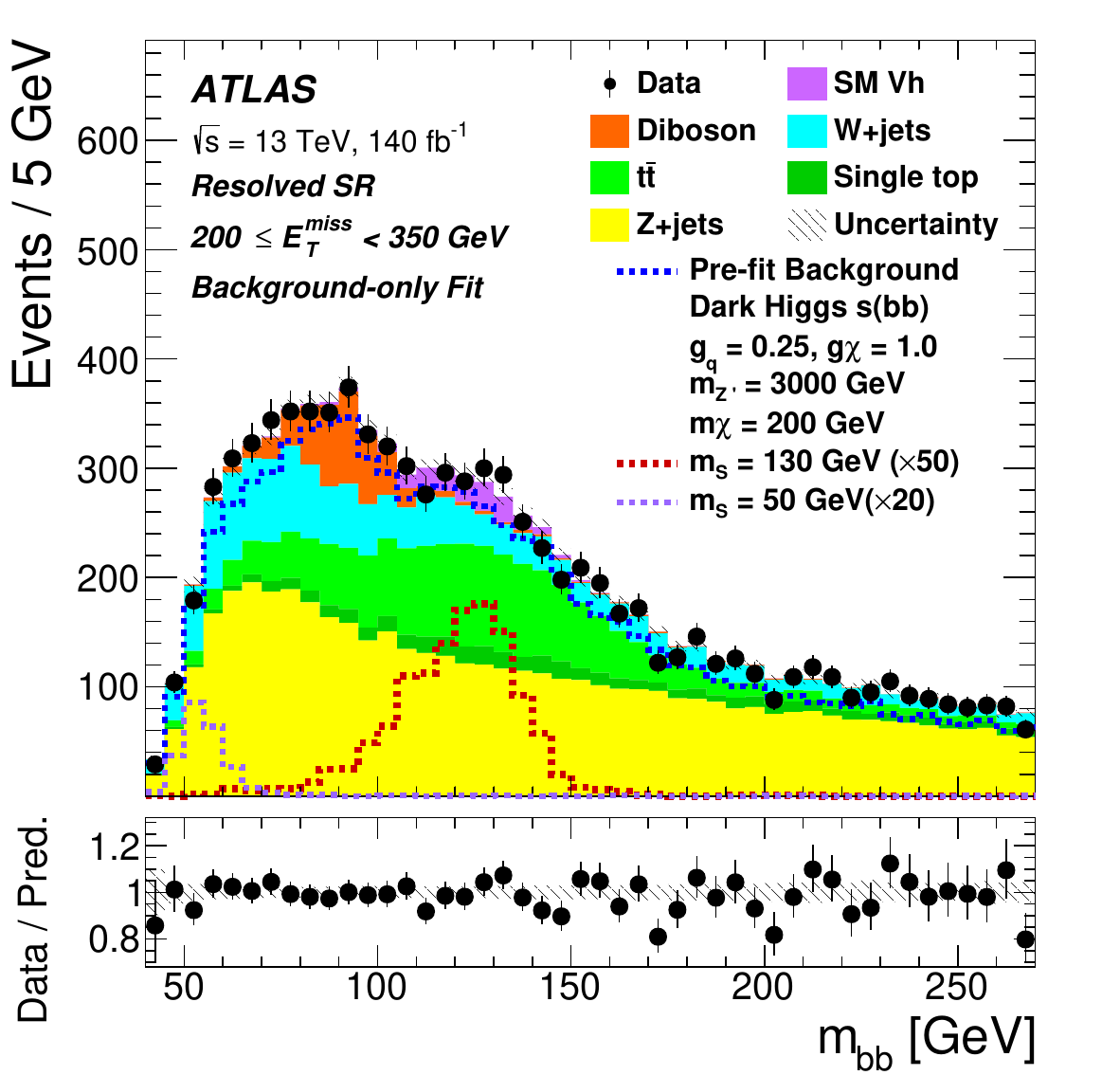} %
\includegraphics[width=0.32\linewidth]{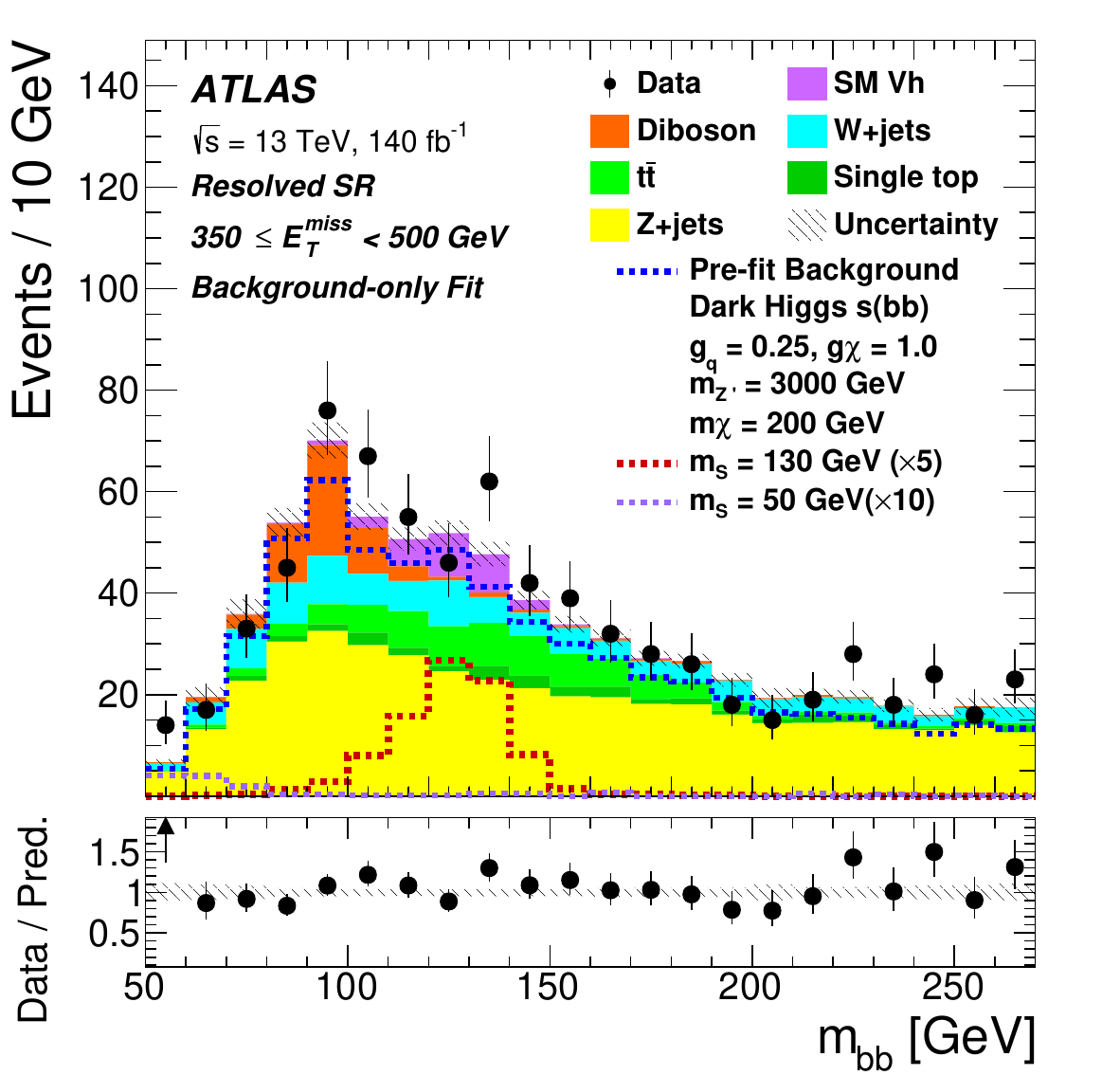} %
\includegraphics[width=0.32\linewidth]{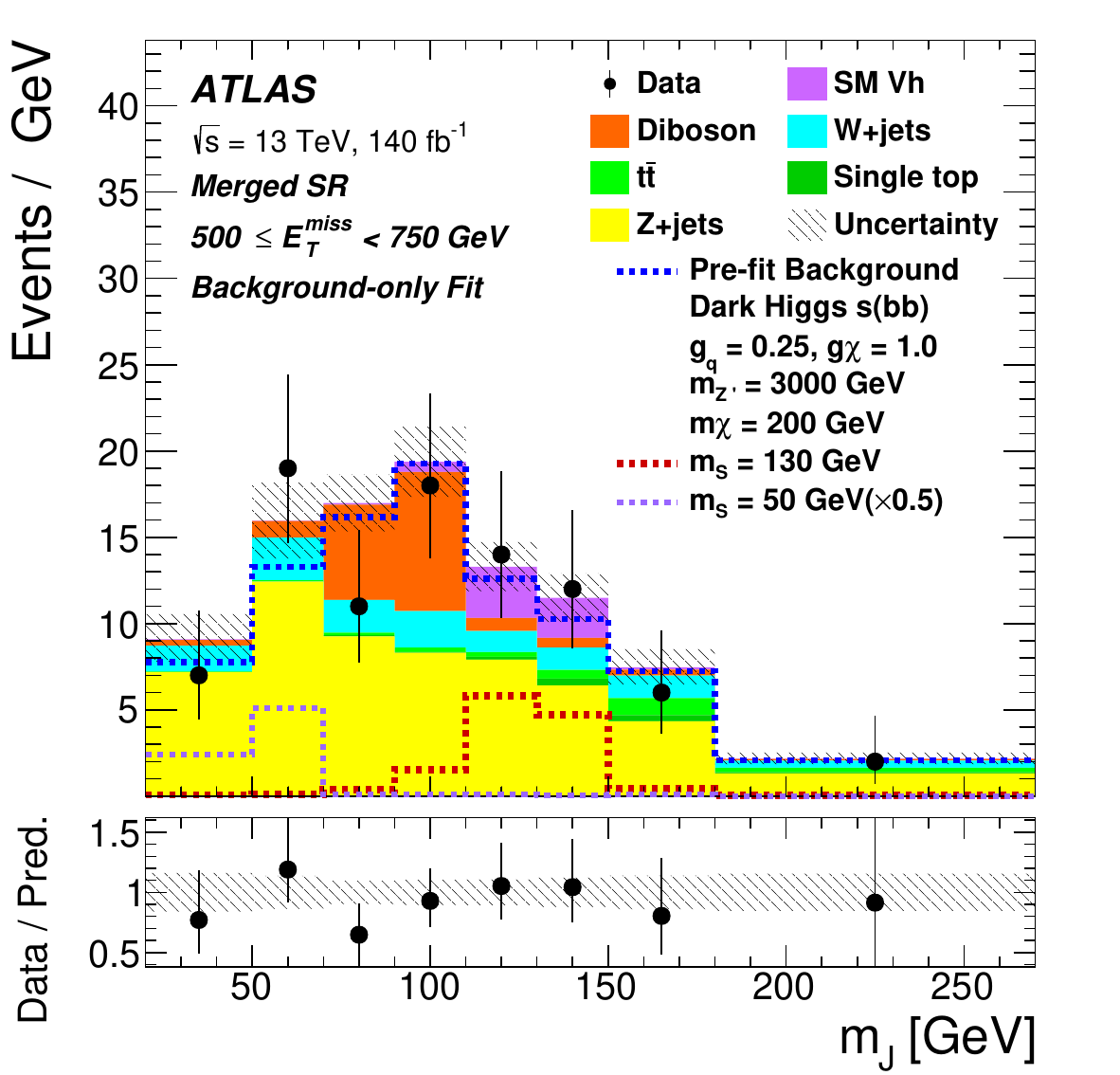} %
\includegraphics[width=0.32\linewidth]{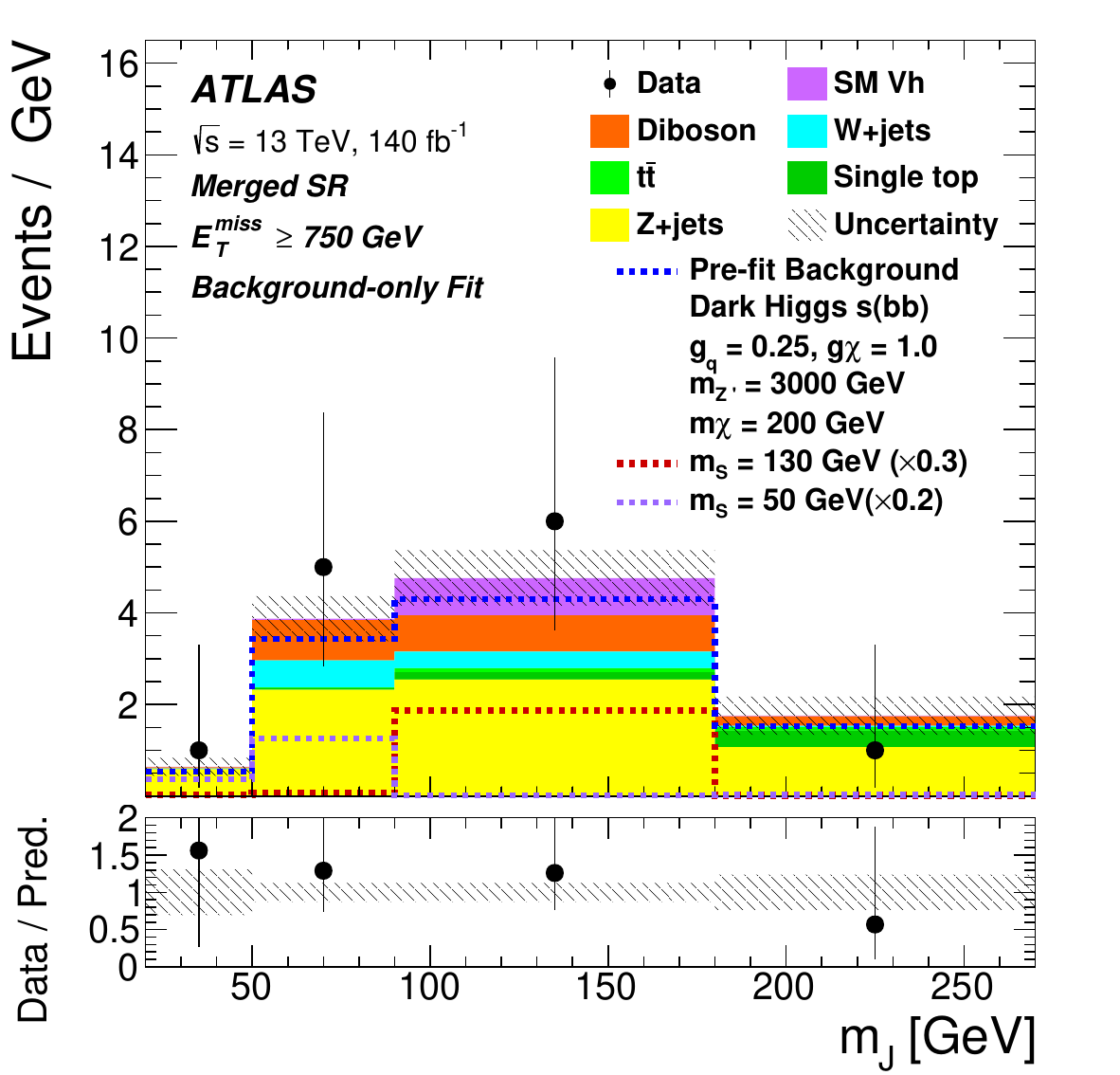} %
\end{center}
\caption{The $m_{bb}$ distributions for data and SM expectations in the different \met~ regions for the resolved (top row) and merged (bottom row) topologies after a background-only simultaneous fit to data.
The shaded area represents the total uncertainty in the predicted yields.
Two signal distributions are overlaid, multiplied in each \met~ region by a scale factor indicated in the legend for visibility. The lower panels show the ratios of the data to the predictions.}
\label{fig:ms-body}
\end{figure}

Consequently, upper limits are set on the product of the $pp\rightarrow s \chi\chi$ production cross-section and branching fraction~$\mathcal{B}(\sbb)$, using a modified frequentist approach~(CL$_\mathrm{s}$)~\cite{Read:2002hq} with a test statistic based on the profile likelihood in the asymptotic approximation~\cite{Cowan:2010js}.
Exclusion contours at $95\%$ confidence level (CL) for the dark Higgs model are presented in Figure~\ref{fig:limit_contour}.
The two-dimensional (\mZp, \ms) plane for scenarios 1 and 2 are shown in Figures~\ref{fig:limit_contour_s1}, \ref{fig:limit_contour_summ} and \ref{fig:limit_contour_s2}.
In scenario 1, with $\gchi=1$ and $\mDM=200$~\GeV, \mZp values are excluded up to $3.4~\TeV$ at $\ms=70~\GeV$, which are the highest mass exclusions for this conventional benchmark model. Figure~\ref{fig:limit_contour_summ} summarizes the exclusions on this model. The observed relic density is obtained for $\mZp = 850~\GeV$, and also for $\ms \simeq 2 \mDM = 400$~\GeV where dark Higgs boson annihilation processes are greatly enhanced and deplete the relic abundance for all $\mZp$ values.

\begin{figure}[!htb]
\centering
\subfloat[]{\includegraphics[width=0.40\textwidth]{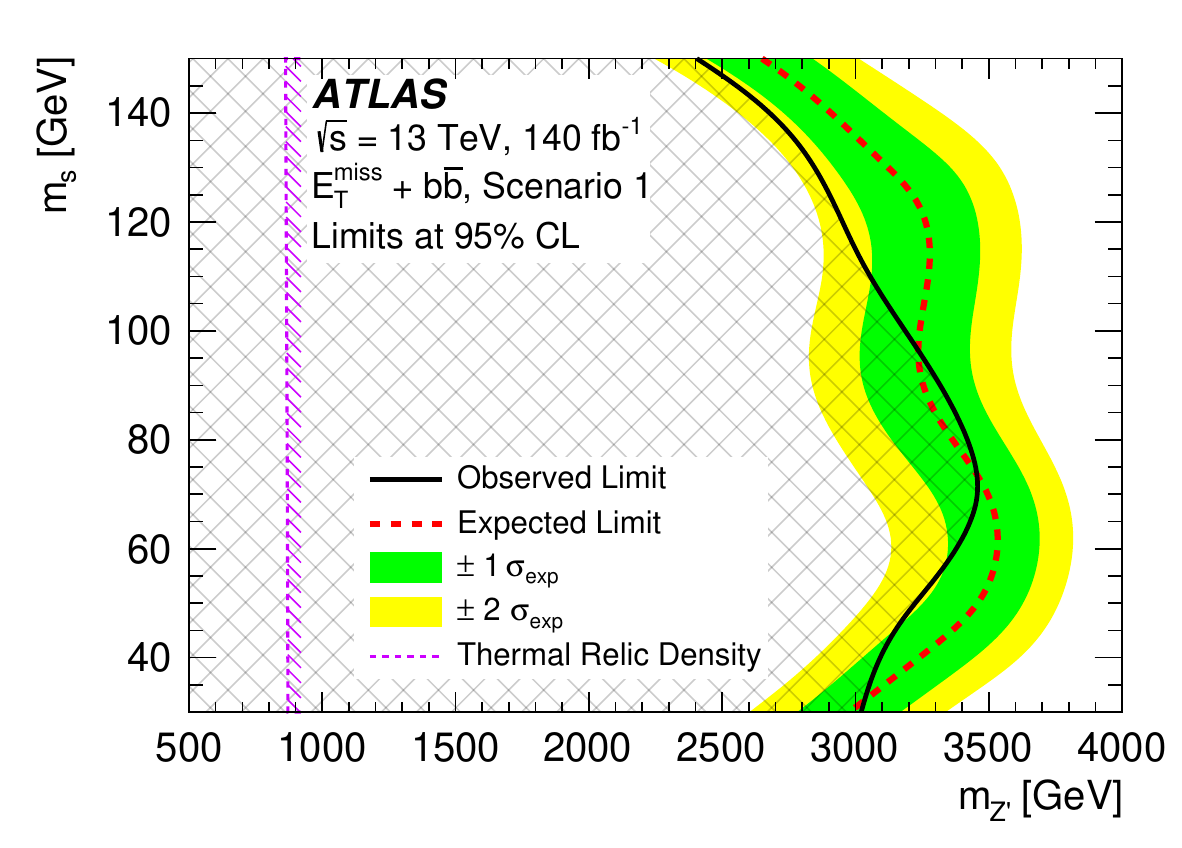} \label{fig:limit_contour_s1}}
\subfloat[]{\includegraphics[width=0.53\textwidth]{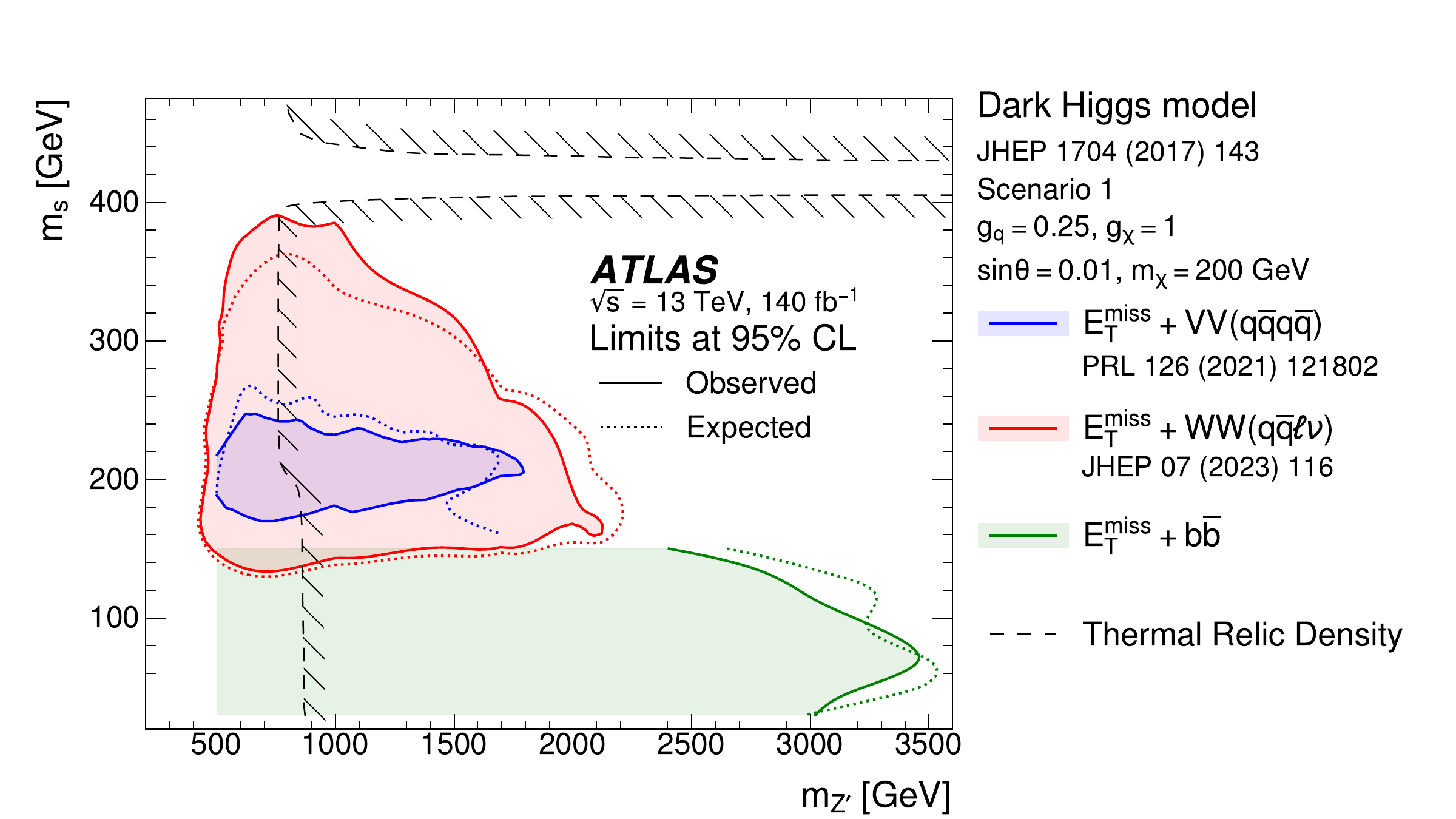} \label{fig:limit_contour_summ}}\\
\subfloat[]{\includegraphics[width=0.40\textwidth]{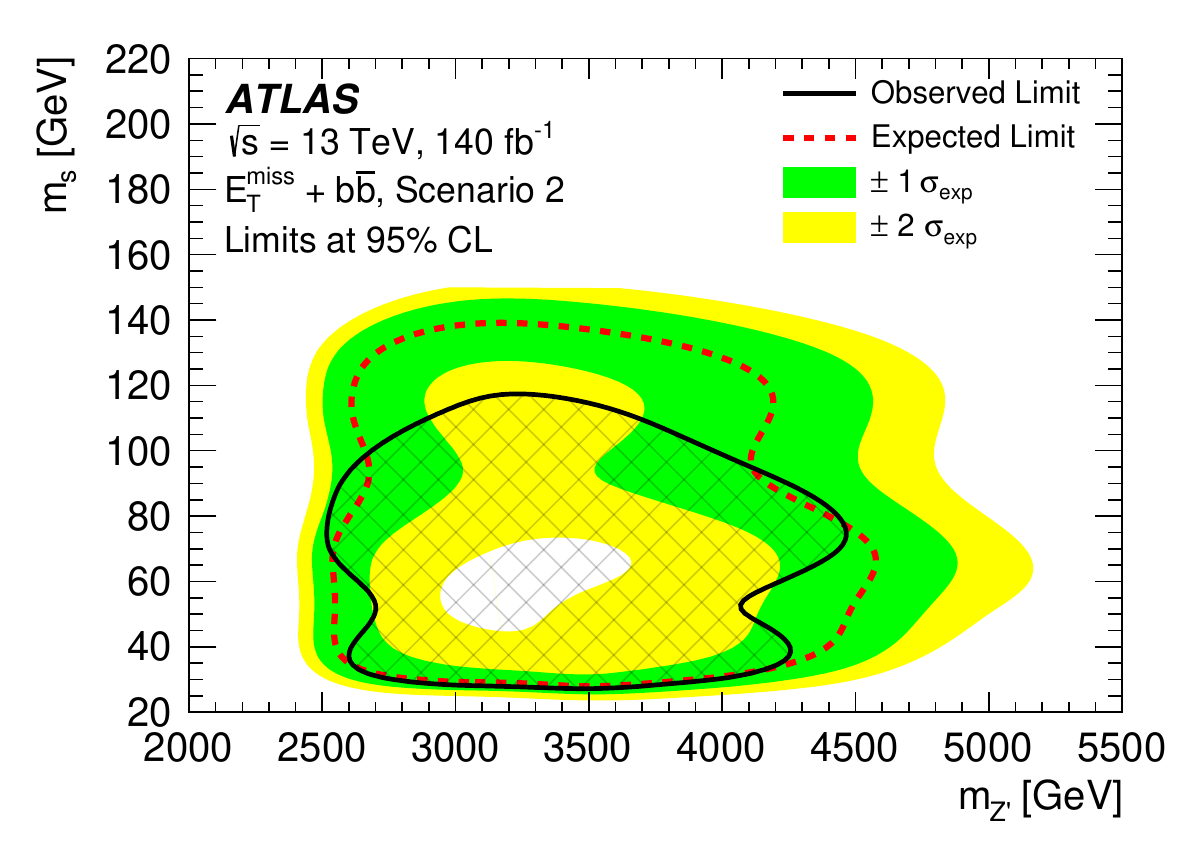} \label{fig:limit_contour_s2}}%
\subfloat[]{\includegraphics[width=0.40\textwidth]{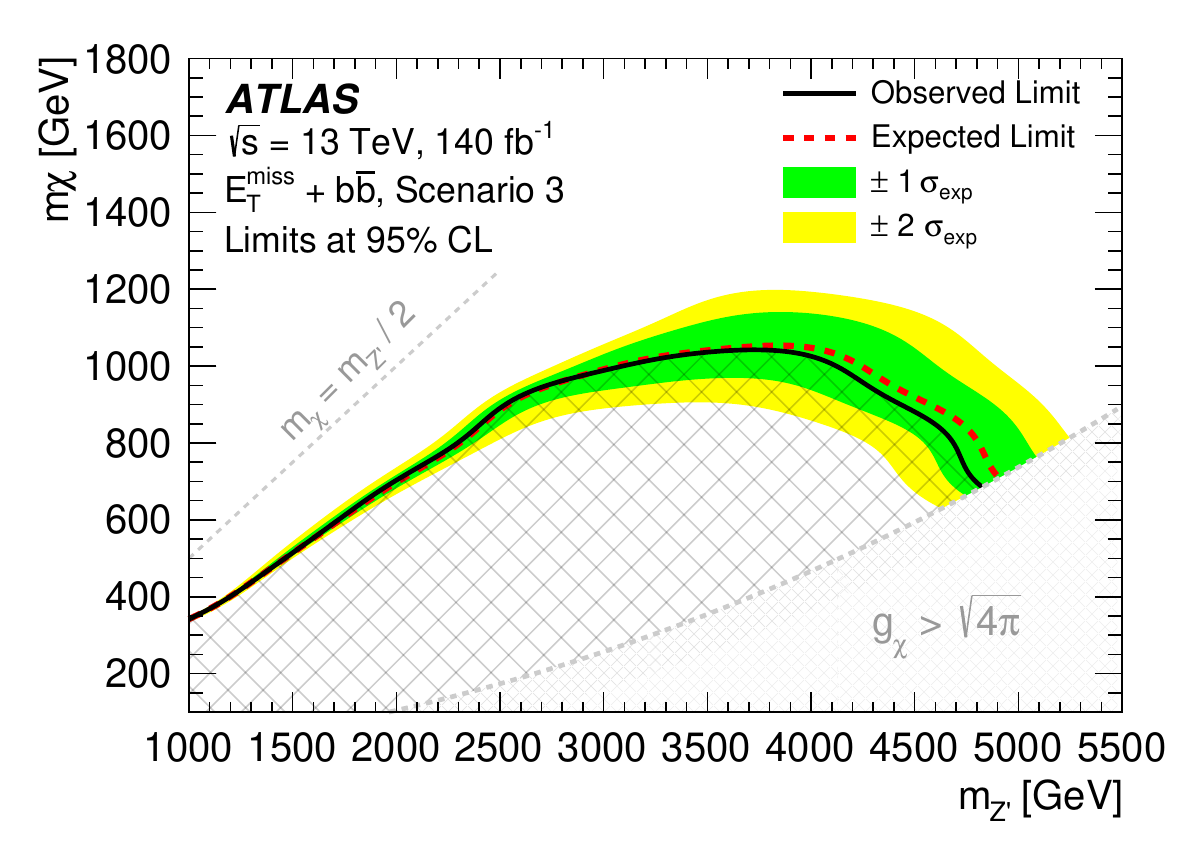} \label{fig:limit_contour_s3}} \\
\caption{
Observed (expected) exclusion regions at 95$\%$ CL for (a, b) scenario 1, (c) scenario 2 and (d) scenario 3.
The $\pm1\sigma$ ($\pm2\sigma$) expected exclusion intervals are shown as the filled inner (outer) bands (a, c, d).
The observed relic density
is indicated with a dashed line
for scenario~1 (a, b), with the diagonal lines indicating an overabundance of DM\@.
The filled areas of (b) are excluded, with the open contours indicating regions not explored for a given signature.
The lower-right shaded area indicates the region beyond the \gchi perturbative limit for scenario~3 (d).
}
\label{fig:limit_contour}
\end{figure}

In scenario 2 (Figure~\ref{fig:limit_contour_s2}), the DM coupling \gchi varies to satisfy the observed relic density throughout; thus the exclusion behavior is more complex.
The increased DM mass ($\mDM=900$~\GeV) leads to reduced cross-sections, with \mZp masses around $\mZp=2.5$~\TeV having \gchi values near unity and lying close to the expected exclusions.
The relic density constraint requires larger $\gchi$ values for larger \mZp values, as the DM annihilation process $\chi\chi \rightarrow \Zp \rightarrow q\bar{q}$ becomes more inefficient. The increasing coupling, and thus greater probability for the \Zp to emit a dark Higgs boson and decay into DM, increases the cross-section, extending sensitivity to higher masses, where \mZp values can be excluded up to $4.5~\TeV$ for $\ms=75~\GeV$. Below $\mZp=2.5$~\TeV, and especially for $\mZp \sim 2 \mDM$, the annihilation process above becomes very efficient, and the small $\gchi$ couplings that match the relic density lead to cross-sections that are too small to be excluded.
The observed exclusion range in \mZp becomes narrower than expected at higher \ms values owing to the small excesses in data near $m_{bb}=50~\GeV$ and $m_{bb}=130~\GeV$ discussed above.

The exclusion limits on the \mZp and $\mDM$ plane for scenario 3 are shown in Figure~\ref{fig:limit_contour_s3}. Again, the relic density depends upon the efficiency of the $\chi\chi \rightarrow \Zp \rightarrow q\bar{q}$ process, resulting in increasing $\gchi$ values for higher \mZp values and lower $\chi$ masses (lower right of figure). For DM masses up to $700$~\GeV, \mZp values up to the perturbative limit are excluded, reaching a maximum of $4.8$~\TeV at that $\mDM$.
The \merged SR dominates the sensitivity at low \ms and high \mZp, while the \resolved SR contributes for $\mZp < 2$~\TeV. %

In conclusion, this Letter reports a novel search for dark matter in a final state with large \met\ and a resonant \bbbar pair with $30 < m_{bb} < 150$~\GeV using \Lumi~\ifb\ of 13~\TeV $pp$ data collected by the ATLAS detector at the LHC\@.
The analysis employs jet reclustering and a new $X\to \bbbar$ tagging algorithm to provide sensitivity to low $m_{bb}$ and highly boosted $bb$-jets.
No excess over the expected background prediction is observed and 95\% CL exclusions are placed on dark Higgs boson models with $m_s<150$~\GeV.
In this dark Higgs boson mass range, $Z'$ mediators are excluded with masses up to 3.4~\TeV, for a benchmark model with $g_\chi=1$, $g_q=0.25$ and $\sin \theta = 0.01$, and up to 4.8~\TeV in a relic density inspired benchmark model in which couplings vary more widely.
This first experimental search for this signature significantly extends existing exclusions on this model, and strongly constrains regions compatible with the observed DM relic density, which occur primarily when the dark Higgs boson is comparable or lighter in mass than the DM candidate, such that related annihilation processes dominate.
These results complement other higher-mass dark Higgs boson and collider DM searches.

%
%
%

%
%

%
%

We thank CERN for the very successful operation of the LHC and its injectors, as well as the support staff at
CERN and at our institutions worldwide without whom ATLAS could not be operated efficiently.

The crucial computing support from all WLCG partners is acknowledged gratefully, in particular from CERN, the ATLAS Tier-1 facilities at TRIUMF/SFU (Canada), NDGF (Denmark, Norway, Sweden), CC-IN2P3 (France), KIT/GridKA (Germany), INFN-CNAF (Italy), NL-T1 (Netherlands), PIC (Spain), RAL (UK) and BNL (USA), the Tier-2 facilities worldwide and large non-WLCG resource providers. Major contributors of computing resources are listed in Ref.~\cite{ATL-SOFT-PUB-2025-001}.

We gratefully acknowledge the support of ANPCyT, Argentina; YerPhI, Armenia; ARC, Australia; BMWFW and FWF, Austria; ANAS, Azerbaijan; CNPq and FAPESP, Brazil; NSERC, NRC and CFI, Canada; CERN; ANID, Chile; CAS, MOST and NSFC, China; Minciencias, Colombia; MEYS CR, Czech Republic; DNRF and DNSRC, Denmark; IN2P3-CNRS and CEA-DRF/IRFU, France; SRNSFG, Georgia; BMBF, HGF and MPG, Germany; GSRI, Greece; RGC and Hong Kong SAR, China; ISF and Benoziyo Center, Israel; INFN, Italy; MEXT and JSPS, Japan; CNRST, Morocco; NWO, Netherlands; RCN, Norway; MNiSW, Poland; FCT, Portugal; MNE/IFA, Romania; MESTD, Serbia; MSSR, Slovakia; ARIS and MVZI, Slovenia; DSI/NRF, South Africa; MICIU/AEI, Spain; SRC and Wallenberg Foundation, Sweden; SERI, SNSF and Cantons of Bern and Geneva, Switzerland; NSTC, Taipei; TENMAK, T\"urkiye; STFC/UKRI, United Kingdom; DOE and NSF, United States of America.

Individual groups and members have received support from BCKDF, CANARIE, CRC and DRAC, Canada; CERN-CZ, FORTE and PRIMUS, Czech Republic; COST, ERC, ERDF, Horizon 2020, ICSC-NextGenerationEU and Marie Sk{\l}odowska-Curie Actions, European Union; Investissements d'Avenir Labex, Investissements d'Avenir Idex and ANR, France; DFG and AvH Foundation, Germany; Herakleitos, Thales and Aristeia programmes co-financed by EU-ESF and the Greek NSRF, Greece; BSF-NSF and MINERVA, Israel; Norwegian Financial Mechanism 2014-2021, Norway; NCN and NAWA, Poland; La Caixa Banking Foundation, CERCA Programme Generalitat de Catalunya and PROMETEO and GenT Programmes Generalitat Valenciana, Spain; G\"{o}ran Gustafssons Stiftelse, Sweden; The Royal Society and Leverhulme Trust, United Kingdom.

In addition, individual members wish to acknowledge support from CERN: European Organization for Nuclear Research (CERN PJAS); Chile: Agencia Nacional de Investigaci\'on y Desarrollo (FONDECYT 1190886, FONDECYT 1230812, FONDECYT 1230987); China: Chinese Ministry of Science and Technology (MOST-2023YFA1605700, MOST-2023YFA1609300), National Natural Science Foundation of China (NSFC - 12175119, NSFC 12275265, NSFC-12075060); Czech Republic: Czech Science Foundation (GACR - 24-11373S), Ministry of Education Youth and Sports (FORTE CZ.02.01.01/00/22\_008/0004632), PRIMUS Research Programme (PRIMUS/21/SCI/017); EU: H2020 European Research Council (ERC - 101002463); European Union: European Research Council (ERC - 948254, ERC 101089007), Horizon 2020 Framework Programme (MUCCA - CHIST-ERA-19-XAI-00), European Union, Future Artificial Intelligence Research (FAIR-NextGenerationEU PE00000013), Italian Center for High Performance Computing, Big Data and Quantum Computing (ICSC, NextGenerationEU); France: Agence Nationale de la Recherche (ANR-20-CE31-0013, ANR-21-CE31-0013, ANR-21-CE31-0022, ANR-22-EDIR-0002), Investissements d'Avenir Labex (ANR-11-LABX-0012); Germany: Baden-Württemberg Stiftung (BW Stiftung-Postdoc Eliteprogramme), Deutsche Forschungsgemeinschaft (DFG - 469666862, DFG - CR 312/5-2); Italy: Istituto Nazionale di Fisica Nucleare (ICSC, NextGenerationEU), Ministero dell'Università e della Ricerca (PRIN - 20223N7F8K - PNRR M4.C2.1.1); Japan: Japan Society for the Promotion of Science (JSPS KAKENHI JP21H05085, JSPS KAKENHI JP22H01227, JSPS KAKENHI JP22H04944, JSPS KAKENHI JP22KK0227); Netherlands: Netherlands Organisation for Scientific Research (NWO Veni 2020 - VI.Veni.202.179); Norway: Research Council of Norway (RCN-314472); Poland: Ministry of Science and Higher Education (IDUB AGH, POB8, D4 no 9722), Polish National Agency for Academic Exchange (PPN/PPO/2020/1/00002/U/00001), Polish National Science Centre (NCN 2021/42/E/ST2/00350, NCN OPUS nr 2022/47/B/ST2/03059, NCN UMO-2019/34/E/ST2/00393, NCN \& H2020 MSCA 945339, UMO-2020/37/B/ST2/01043, UMO-2021/40/C/ST2/00187, UMO-2022/47/O/ST2/00148, UMO-2023/49/B/ST2/04085); Slovenia: Slovenian Research Agency (ARIS grant J1-3010); Spain: Generalitat Valenciana (Artemisa, FEDER, IDIFEDER/2018/048), Ministry of Science and Innovation (MCIN \& NextGenEU PCI2022-135018-2, MICIN \& FEDER PID2021-125273NB, RYC2019-028510-I, RYC2020-030254-I, RYC2021-031273-I, RYC2022-038164-I), PROMETEO and GenT Programmes Generalitat Valenciana (CIDEGENT/2019/023, CIDEGENT/2019/027); Sweden: Carl Trygger Foundation (Carl Trygger Foundation CTS 22:2312), Swedish Research Council (Swedish Research Council 2023-04654, VR 2018-00482, VR 2022-03845, VR 2022-04683, VR 2023-03403, VR grant 2021-03651), Knut and Alice Wallenberg Foundation (KAW 2018.0157, KAW 2018.0458, KAW 2019.0447, KAW 2022.0358); Switzerland: Swiss National Science Foundation (SNSF - PCEFP2\_194658); United Kingdom: Leverhulme Trust (Leverhulme Trust RPG-2020-004), Royal Society (NIF-R1-231091); United States of America: U.S. Department of Energy (ECA DE-AC02-76SF00515), Neubauer Family Foundation.

%
%


%
%
%
%
%
%
%

%
%
\appendix
\part*{Appendix}
\addcontentsline{toc}{part}{Appendix}

\subsubsection*{The ATLAS detector and trigger system}

The ATLAS experiment is a multipurpose particle detector with a forward--backward symmetric cylindrical geometry and nearly $4\pi$ coverage in solid angle. It consists of an inner tracking detector surrounded by a superconducting solenoid, sampling electromagnetic (EM) and hadronic calorimeters, and a muon spectrometer with three toroidal superconducting magnets.
A two-level trigger system~\cite{TRIG-2016-01} selects events for offline analysis. Events in the SR and the single-muon CR were collected by triggers on the \met reconstructed from calorimeter information only~\cite{TRIG-2019-01} above a threshold that varied from $90$ to $110~\GeV$.
Events in the two-lepton CR were recorded using single-lepton triggers with $\pt$ thresholds of $24$--$26~\GeV$~\cite{TRIG-2018-05,TRIG-2018-01}, depending on the data-taking period.
An extensive software suite~\cite{SOFT-2022-02} is used in
the experiment.

\subsubsection*{Simulation of signal and background processes}

\begin{figure}[!t]
\begin{center}
\includegraphics[width=0.25\linewidth]{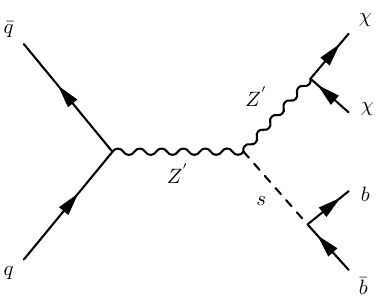} \hspace{1.2cm}%
\includegraphics[width=0.25\linewidth]{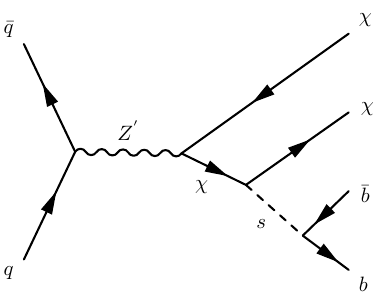}
\end{center}
\caption{Signal diagrams illustrating the resonant $pp \rightarrow \Zprime \rightarrow s\chi\chi \rightarrow b\bar{b}\chi\chi$ process. }
\label{fig:feynman}
\end{figure}

Simulated signal samples for the $pp \rightarrow \Zprime \rightarrow s\chi\chi \rightarrow b\bar{b}\chi\chi$ process, illustrated in Figure~\ref{fig:feynman},
were generated at LO in quantum chromodynamics (QCD) with up to one additional parton in the event, using \MGNLO[2.9.3]~\cite{Alwall:2014hca} interfaced to \PYTHIA[8.245]~\cite{Sjostrand:2014zea}, both using the NNPDF3.0~LO PDF set~\cite{Ball:2014uwa} with $\alphas=0.13$~\cite{Ball:2014uwa} and the A14 set of tuned parameters~\cite{ATL-PHYS-PUB-2014-021}.

The \vjets background was simulated with \SHERPA[2.2.11]~\cite{Bothmann:2019yzt}, using next-to-leading-order (NLO) matrix elements for up to two partons, and LO matrix elements for up to five partons calculated with the \textsc{Comix}~\cite{Gleisberg:2008fv} and \OPENLOOPS~\cite{Buccioni:2019sur,Cascioli:2011va,Denner:2016kdg} libraries. The matching to the \SHERPA parton shower~\cite{Schumann:2007mg} used the \MEPSatNLO prescription~\cite{Hoeche:2011fd,Hoeche:2012yf,Catani:2001cc,Hoeche:2009rj} with the set of tuned parameters developed by the \SHERPA authors. The \NNPDF[3.0]~NNLO set of PDFs~\cite{Ball:2014uwa} was used and the samples were normalized to the next-to-next-to-leading-order (NNLO) prediction~\cite{Anastasiou:2003ds}.
Backgrounds from \ttbar production and single top quark production were generated at NLO in QCD with \POWHEGBOX[v2]~\cite{Frixione:2007nw,Nason:2004rx,Frixione:2007vw,Alioli:2010xd} using the NNPDF3.0~NLO PDF set, interfaced to \PYTHIA[8.230]. Parton shower simulations with \PYTHIA[8.230] used the A14 set of tuned parameters~\cite{ATL-PHYS-PUB-2014-021} with the NNPDF2.3~LO PDF set.
The \ttbar samples were normalized using calculations at NNLO in QCD including next-to-next-to-leading logarithmic soft-gluon terms calculated using \TOPpp[2.0]~\cite{Beneke:2011mq,Cacciari:2011hy,Baernreuther:2012ws,Czakon:2012zr,Czakon:2012pz,Czakon:2013goa,Czakon:2011xx}. The single-top-quark processes were normalized to cross-sections at \NLO in QCD from \HATHOR[v2.1]~\cite{Aliev:2010zk,Kant:2014oha}.
Samples of diboson final states (\(VV\)) were simulated with the
\SHERPA[2.2.1] or 2.2.2~\cite{Bothmann:2019yzt} generator depending on the process and normalized using calculations at NNLO in QCD using the NNPDF3.0~NNLO PDF set.
Backgrounds from associated $Vh$ production were generated at \NLO in QCD with \POWHEGBOX interfaced to \PYTHIA[8.186] using the NNPDF3.0~NLO PDF set. The $qq \to Vh$ and $gg \to Vh$ processes were normalized using calculations at NNLO in QCD and at NLO in QCD combined with next-to-leading-logarithmic order corrections, respectively~\cite{Ciccolini:2003jy,Brein:2003wg,Brein:2011vx,Altenkamp:2012sx,Denner:2014cla,Brein:2012ne,Harlander:2014wda}.
Top quarks were decayed at LO using \MADSPIN~\cite{Frixione:2007zp,Artoisenet:2012st} to preserve all spin correlations.
The decays of bottom and charm hadrons were simulated
using the \EVTGEN[1.6.0] program~\cite{Lange:2001uf}.
For all signal and background simulations, contributions from additional \pp interactions in the same and neighboring bunch crossings (pileup) were simulated through the overlay of inelastic \pp simulations from \PYTHIA[8.186]~\cite{Sjostrand:2007gs} using the A3 set of tuned parameters~\cite{ATL-PHYS-PUB-2012-003} and the NNPDF2.3 leading order (LO) parton distribution function (PDF) set~\cite{Ball:2012cx}.

Theoretical and modelling uncertainties assessed include those from the choice of PDFs and the factorization and renormalization scales. Additionally, uncertainties in the choice of the matrix element and parton shower generator are assessed through dedicated, alternative MC simulations.
For top quark processes, uncertainties in the choice of generator were evaluated by comparison with event samples generated with \MGNLO[2.6.0] interfaced to \PYTHIA[8.230] and the nominal \POWHEG generator hadronized by \HERWIG[7.04]~\cite{Bahr:2008pv,Bellm:2015jjp}, using the H7UE set of tuned parameters~\cite{Bellm:2015jjp} and the \MMHT\ LO PDF set~\cite{Harland-Lang:2014zoa}. For single top quark production in the $tW$ channel, an alternative sample was generated using the diagram subtraction scheme~\cite{Frixione:2008yi,ATL-PHYS-PUB-2016-020} to estimate the uncertainty arising from the interference with \ttbar production.
For \vjets processes, a sample generated with \MGNLO[2.6.2] at LO in QCD with up to four parton emissions using the NNPDF2.3~LO PDF set and interfaced to \PYTHIA[8.230] using a merging scale of $Q_\text{cut} = 30~\GeV$ was employed. Uncertainties in the matching parameter and resummation scale were also assessed.

\printbibliography
\clearpage
\input{atlas_authlist}
%

%
%

%

%
%
%
%

\end{document}

%% file: atlas_authlist.tex
 
\begin{flushleft}
\hypersetup{urlcolor=black}
{\Large The ATLAS Collaboration}

\bigskip

\AtlasOrcid[0000-0002-6665-4934]{G.~Aad}$^\textrm{\scriptsize 104}$,
\AtlasOrcid[0000-0001-7616-1554]{E.~Aakvaag}$^\textrm{\scriptsize 17}$,
\AtlasOrcid[0000-0002-5888-2734]{B.~Abbott}$^\textrm{\scriptsize 123}$,
\AtlasOrcid[0000-0002-0287-5869]{S.~Abdelhameed}$^\textrm{\scriptsize 119a}$,
\AtlasOrcid[0000-0002-1002-1652]{K.~Abeling}$^\textrm{\scriptsize 56}$,
\AtlasOrcid[0000-0001-5763-2760]{N.J.~Abicht}$^\textrm{\scriptsize 50}$,
\AtlasOrcid[0000-0002-8496-9294]{S.H.~Abidi}$^\textrm{\scriptsize 30}$,
\AtlasOrcid[0009-0003-6578-220X]{M.~Aboelela}$^\textrm{\scriptsize 45}$,
\AtlasOrcid[0000-0002-9987-2292]{A.~Aboulhorma}$^\textrm{\scriptsize 36e}$,
\AtlasOrcid[0000-0001-5329-6640]{H.~Abramowicz}$^\textrm{\scriptsize 155}$
% ...
\bigskip
\\

% ...
$^{17}$Department for Physics and Technology, University of Bergen, Bergen; Norway.\\
% ...
$^{30}$Physics Department, Brookhaven National Laboratory, Upton NY; United States of America.\\
% ...
$^{36}$$^{(a)}$Facult\'e des Sciences Ain Chock, Universit\'e Hassan II de Casablanca;$^{(b)}$Facult\'{e} des Sciences, Universit\'{e} Ibn-Tofail, K\'{e}nitra;$^{(c)}$Facult\'e des Sciences Semlalia, Universit\'e Cadi Ayyad, LPHEA-Marrakech;$^{(d)}$LPMR, Facult\'e des Sciences, Universit\'e Mohamed Premier, Oujda;$^{(e)}$Facult\'e des sciences, Universit\'e Mohammed V, Rabat;$^{(f)}$Institute of Applied Physics, Mohammed VI Polytechnic University, Ben Guerir; Morocco.\\
% ...
$^{45}$Physics Department, Southern Methodist University, Dallas TX; United States of America.\\
% ...
$^{50}$Fakult\"{a}t Physik , Technische Universit{\"a}t Dortmund, Dortmund; Germany.\\
% ...
$^{56}$II. Physikalisches Institut, Georg-August-Universit\"{a}t G\"ottingen, G\"ottingen; Germany.\\
% ...
$^{104}$CPPM, Aix-Marseille Universit\'e, CNRS/IN2P3, Marseille; France.\\
% ...
$^{119}$$^{(a)}$New York University Abu Dhabi, Abu Dhabi;$^{(b)}$United Arab Emirates University, Al Ain; United Arab Emirates.\\
% ...
$^{123}$Homer L. Dodge Department of Physics and Astronomy, University of Oklahoma, Norman OK; United States of America.\\
% ...
$^{155}$Raymond and Beverly Sackler School of Physics and Astronomy, Tel Aviv University, Tel Aviv; Israel.\\
% ...
$^{a}$ Also Affiliated with an institute covered by a cooperation agreement with CERN.\\
$^{b}$ Also at An-Najah National University, Nablus; Palestine.\\
$^{c}$ Also at Borough of Manhattan Community College, City University of New York, New York NY; United States of America.\\
$^{d}$ Also at Center for High Energy Physics, Peking University; China.\\
$^{e}$ Also at Center for Interdisciplinary Research and Innovation (CIRI-AUTH), Thessaloniki; Greece.\\
$^{f}$ Also at Centro Studi e Ricerche Enrico Fermi; Italy.\\
% ...

\end{flushleft}


%% file: ANA-EXOT-2020-26-PAPER.bib
@book{Bertone:2010zza,
    author = "Silk, J. and others",
    editor = "Bertone, Gianfranco",
    title = "{Particle Dark Matter: Observations, Models and Searches}",
    doi = "10.1017/CBO9780511770739",
    isbn = "978-1-107-65392-4",
    publisher = "Cambridge Univ. Press",
    address = "Cambridge",
    year = "2010"
}

@article{Feng:2010gw,
    author = "Feng, Jonathan L.",
    title = "{Dark Matter Candidates from Particle Physics and Methods of Detection}",
    eprint = "1003.0904",
    archivePrefix = "arXiv",
    primaryClass = "astro-ph.CO",
    reportNumber = "UCI-TR-2009-13",
    doi = "10.1146/annurev-astro-082708-101659",
    journal = "Ann. Rev. Astron. Astrophys.",
    volume = "48",
    pages = "495--545",
    year = "2010"
}

@article{Porter:2011nv,
    author = "Porter, Troy A. and Johnson, Robert P. and Graham, Peter W.",
    title = "{Dark Matter Searches with Astroparticle Data}",
    eprint = "1104.2836",
    archivePrefix = "arXiv",
    primaryClass = "astro-ph.HE",
    doi = "10.1146/annurev-astro-081710-102528",
    journal = "Ann. Rev. Astron. Astrophys.",
    volume = "49",
    pages = "155--194",
    year = "2011"
}

@article{Bertone:2017adx,
    author = "Bertone, Gianfranco and Bozorgnia, Nassim and Kim, Jong Soo and Liem, Sebastian and McCabe, Christopher and Otten, Sydney and Ruiz de Austri, Roberto",
    title = "{Identifying WIMP dark matter from particle and astroparticle data}",
    eprint = "1712.04793",
    archivePrefix = "arXiv",
    primaryClass = "hep-ph",
    reportNumber = "IPPP-17-94, KCL-PH-TH-2017-55, TTK-17-45",
    doi = "10.1088/1475-7516/2018/03/026",
    journal = "JCAP",
    volume = "03",
    pages = "026",
    year = "2018"
}

@article{PhysRevLett.121.111302,
  title     = {Dark Matter Search Results from a One Ton-Year Exposure of XENON1T},
  author    = {{XENON Collaboration}},
  eprint = "1805.12562",
  archivePrefix = "arXiv",
  primaryClass = "astro-ph.CO",
  journal   = {Phys. Rev. Lett.},
  volume    = {121},
  pages     = {111302},
  numpages  = {8},
  year      = {2018},
  month     = {Sep},
  publisher = {American Physical Society},
  doi       = {10.1103/PhysRevLett.121.111302}
}

@article{PhysRevLett.117.121303,
  title     = {Dark Matter Results from First 98.7 Days of Data from the PandaX-II Experiment},
  author    = {{PandaX-II Collaboration}},
  eprint = "1607.07400",
  archivePrefix = "arXiv",
  primaryClass = "hep-ex",
  journal   = {Phys. Rev. Lett.},
  volume    = {117},
  pages     = {121303},
  numpages  = {7},
  year      = {2016},
  month     = {Sep},
  publisher = {American Physical Society},
  doi       = {10.1103/PhysRevLett.117.121303}
}

@article{PhysRevLett.121.081307,
  title     = {Low-Mass Dark Matter Search with the DarkSide-50 Experiment},
  author    = {{DarkSide Collaboration}},
  eprint    = "1802.06994",
  archivePrefix = "arXiv",
  primaryClass = "astro.ph.HE",
  journal   = {Phys. Rev. Lett.},
  volume    = {121},
  pages     = {081307},
  numpages  = {8},
  year      = {2018},
  month     = {Aug},
  publisher = {American Physical Society},
  doi       = {10.1103/PhysRevLett.121.081307}
}

@article{PhysRevLett.123.241803,
  title     = {Search for Light Dark Matter Interactions Enhanced by the Migdal Effect or Bremsstrahlung in XENON1T},
  author    = {{XENON Collaboration}},
  eprint    = "1907.12771",
  archivePrefix = "arXiv",
  primaryClass = "hep-ex",
  journal   = {Phys. Rev. Lett.},
  volume    = {123},
  pages     = {241803},
  numpages  = {7},
  year      = {2019},
  month     = {Dec},
  publisher = {American Physical Society},
  doi       = {10.1103/PhysRevLett.123.241803}
}

@article{PhysRevD.100.022001,
  title     = {Dark matter search results from the complete exposure of the PICO-60 ${\mathrm{C}}_{3}{\mathrm{F}}_{8}$ bubble chamber},
  author    = {{PICO Collaboration}},
  eprint    = "1902.04031",
  archivePrefix = "arXiv",
  primaryClass = "astro-ph.CO",
  journal   = {Phys. Rev. D},
  volume    = {100},
  pages     = {022001},
  numpages  = {9},
  year      = {2019},
  month     = {Jul},
  publisher = {American Physical Society},
  doi       = {10.1103/PhysRevD.100.022001}
}

@article{Ackermann_2015,
  title     = {Searching for Dark Matter Annihilation from Milky Way Dwarf Spheroidal Galaxies with Six Years of Fermi Large Area Telescope Data},
  eprint    = "1503.02641",
  archivePrefix = "arXiv",
  primaryClass = "astro-ph.HE",
  volume    = {115},
  pages     = {231301},
  doi       = {10.1103/physrevlett.115.231301},
  number    = {23},
  journal   = {Phys. Rev. Lett.},
  publisher = {American Physical Society (APS)},
  author    = {{Fermi-LAT Collaboration}},
  year      = {2015},
  month     = {Nov}
}

@article{PerezdelosHeros:2020qyt,
    author = "P\'erez de los Heros, Carlos",
    title = "{Status, Challenges and Directions in Indirect Dark Matter Searches}",
    eprint = "2008.11561",
    archivePrefix = "arXiv",
    primaryClass = "astro-ph.HE",
    doi = "10.3390/sym12101648",
    journal = "Symmetry",
    volume = "12",
    number = "10",
    pages = "1648",
    year = "2020"
}

@article{Duerr:2017uap,
    author = "Duerr, Michael and Grohsjean, Alexander and Kahlhoefer, Felix and Penning, Bjoern and Schmidt-Hoberg, Kai and Schwanenberger, Christian",
    title = "{Hunting the dark Higgs}",
    eprint = "1701.08780",
    archivePrefix = "arXiv",
    primaryClass = "hep-ph",
    reportNumber = "DESY-17-016",
    doi = "10.1007/JHEP04(2017)143",
    journal = "JHEP",
    volume = "04",
    pages = "143",
    year = "2017"
}

@article{Abercrombie:2015wmb,
    author = "Abercrombie, Daniel and others",
    %editor = "Boveia, Antonio and Doglioni, Caterina and Lowette, Steven and Malik, Sarah and Mrenna, Stephen",
    title = "{Dark Matter benchmark models for early LHC Run-2 Searches: Report of the ATLAS/CMS Dark Matter Forum}",
    eprint = "1507.00966",
    archivePrefix = "arXiv",
    primaryClass = "hep-ex",
    reportNumber = "FERMILAB-PUB-15-282-CD",
    doi = "10.1016/j.dark.2019.100371",
    journal = "Phys. Dark Univ.",
    volume = "27",
    pages = "100371",
    year = "2020"
}

@article{Abdallah:2015ter,
    author = "Abdallah, Jalal and others",
    title = "{Simplified models for dark matter searches at the LHC}",
    eprint = "1506.03116",
    archivePrefix = "arXiv",
    primaryClass = "hep-ph",
    reportNumber = "FERMILAB-PUB-15-283-CD, CERN-PH-TH-2015-139",
    doi = "10.1016/j.dark.2015.08.001",
    journal = "Phys. Dark Univ.",
    volume = "9-10",
    pages = "8--23",
    year = "2015"
}

@article{MadDM,
  title     = {MadDM v.3.0: A comprehensive tool for dark matter studies},
  eprint = "1804.00044",
  archivePrefix = "arXiv",
  primaryClass = "hep-ph",
  volume    = {24},
  %issn      = {2212-6864},
  %url       = {http://dx.doi.org/10.1016/j.dark.2018.11.009},
  doi       = {10.1016/j.dark.2018.11.009},
  journal   = {Phys. Dark Univ.},
  publisher = {Elsevier BV},
  author    = {Ambrogi, Federico and Arina, Chiara and Backović, Mihailo and Heisig, Jan and Maltoni, Fabio and Mantani, Luca and Mattelaer, Olivier and Mohlabeng, Gopolang},
  year      = {2019},
  month     = {03},
  pages     = {100249}
}

@article{Cacciari:2008gn,
    author = "Cacciari, Matteo and Salam, Gavin P. and Soyez, Gregory",
    title = "{The catchment area of jets}",
    eprint = "0802.1188",
    archivePrefix = "arXiv",
    primaryClass = "hep-ph",
    reportNumber = "LPTHE-07-02",
    doi = "10.1088/1126-6708/2008/04/005",
    journal = "JHEP",
    volume = "04",
    pages = "005",
    year = "2008"
}

@article{Krohn:2009th,
      author         = "Krohn, David and Thaler, Jesse and Wang, Lian-Tao",
      title          = "{Jet Trimming}",
      journal        = "JHEP",
      volume         = "02",
      year           = "2010",
      pages          = "084",
      doi            = "10.1007/JHEP02(2010)084",
      eprint         = "0912.1342",
      archivePrefix  = "arXiv",
      primaryClass   = "hep-ph",
      SLACcitation   = "%%CITATION = ARXIV:0912.1342;%%"
}

@article{Moneta:2010pm,
      author         = "Moneta, Lorenzo and Belasco, Kevin and Cranmer, Kyle S.
                        and Kreiss, S. and Lazzaro, Alfio and Piparo, Danilo and
                        Schott, Gregory and Verkerke, Wouter and Wolf, Matthias",
      title          = "{The RooStats Project}",
      booktitleNOT      = "{Proceedings, 13th International Workshop on Advanced
                        computing and analysis techniques in physics research
                        (ACAT2010)}",
      journal        = "PoS ACAT2010",
      pages          = "057",
      year           = "2010",
      eprint         = "1009.1003",
      archivePrefix  = "arXiv",
      primaryClass   = "physics.data-an",
      SLACcitation   = "%%CITATION = ARXIV:1009.1003;%%"
}

@article{Planck:2018vyg,
    %author = "Aghanim, N. and others",
    author = "{Planck Collaboration}",
    title = "{Planck 2018 results. VI. Cosmological parameters}",
    eprint = "1807.06209",
    archivePrefix = "arXiv",
    primaryClass = "astro-ph.CO",
    doi = "10.1051/0004-6361/201833910",
    journal = "Astron. Astrophys.",
    volume = "641",
    pages = "A6",
    year = "2020",
    %note = "[Erratum: Astron.Astrophys. 652 (2021) C4]"
    related = "Planck:2018vyg-err",
    relatedstring = "Errantum:",
}


%% file: ATLAS-useful.bib
@Article{Cacciari:2008gp,
     author    = "Cacciari, Matteo and Salam, Gavin P. and Soyez, Gregory",
     title     = "{The anti-\(k_{t}\) jet clustering algorithm}",
     journal   = "JHEP",
     volume    = "04",
     year      = "2008",
     pages     = "063",
     eprint    = "0802.1189",
     archivePrefix = "arXiv",
     primaryClass  =  "hep-ph",
     doi       = "10.1088/1126-6708/2008/04/063",
     SLACcitation  = "%%CITATION = 0802.1189;%%"
}

@Article{Fastjet,
      author         = "Cacciari, Matteo and Salam, Gavin P. and Soyez, Gregory",
      title          = "{FastJet user manual}",
      journal        = "Eur. Phys. J. C",
      volume         = "72",
      year           = "2012",
      pages          = "1896",
      doi            = "10.1140/epjc/s10052-012-1896-2",
      eprint         = "1111.6097",
      archivePrefix  = "arXiv",
      primaryClass   = "hep-ph",
      reportNumber   = "CERN-PH-TH-2011-297",
      SLACcitation   = "%%CITATION = ARXIV:1111.6097;%%"
}

@article{Harland-Lang:2014zoa,
      author         = "Harland-Lang, L. A. and Martin, A. D. and Motylinski, P.
                        and Thorne, R. S.",
      title          = "{Parton distributions in the LHC era: MMHT 2014 PDFs}",
      journal        = "Eur. Phys. J. C",
      number         = "5",
      volume         = "75",
      pages          = "204",
      doi            = "10.1140/epjc/s10052-015-3397-6",
      year           = "2015",
      eprint         = "1412.3989",
      archivePrefix  = "arXiv",
      primaryClass   = "hep-ph",
      reportNumber   = "LCTS-2014-47, IPPP-14-97, DCPT-14-194",
      SLACcitation   = "%%CITATION = ARXIV:1412.3989;%%",
}

@article{Ball:2012cx,
      author         = "{NNPDF Collaboration} and Ball, Richard D. and others",
      title          = "{Parton distributions with LHC data}",
      journal        = "Nucl. Phys. B",
      volume         = "867",
      year           = "2013",
      pages          = "244",
      doi            = "10.1016/j.nuclphysb.2012.10.003",
      eprint         = "1207.1303",
      archivePrefix  = "arXiv",
      primaryClass   = "hep-ph",
      reportNumber   = "EDINBURGH-2012-08, IFUM-FT-997, FR-PHENO-2012-014,
                        RWTH-TTK-12-25, CERN-PH-TH-2012-037, SFB-CPP-12-47\,
                        --CERN-PH-TH-2012-037",
      SLACcitation   = "%%CITATION = ARXIV:1207.1303;%%"
}

@article{Ball:2014uwa,
      author         = "{NNPDF Collaboration} and Ball, Richard D. and others",
      title          = "{Parton distributions for the LHC run II}",
      journal        = "JHEP",
      volume         = "04",
      year           = "2015",
      pages          = "040",
      doi            = "10.1007/JHEP04(2015)040",
      eprint         = "1410.8849",
      archivePrefix  = "arXiv",
      primaryClass   = "hep-ph",
      reportNumber   = "EDINBURGH-2014-15, IFUM-1034-FT, CERN-PH-TH-2013-253,
                        OUTP-14-11P, CAVENDISH-HEP-14-11",
      SLACcitation   = "%%CITATION = ARXIV:1410.8849;%%"
}

@article{Czakon:2012pz,
      author         = "Czakon, Michal and Mitov, Alexander",
      title          = "{NNLO corrections to top pair production at hadron
                        colliders: the quark-gluon reaction}",
      journal        = "JHEP",
      volume         = "01",
      pages          = "080",
      doi            = "10.1007/JHEP01(2013)080",
      year           = "2013",
      eprint         = "1210.6832",
      archivePrefix  = "arXiv",
      primaryClass   = "hep-ph",
      SLACcitation   = "%%CITATION = ARXIV:1210.6832;%%",
}

@Article{Sjostrand:2007gs,
	author    = "Sj{\"o}strand, T. and Mrenna, S. and Skands, P.",
	title     = "{A brief introduction to PYTHIA 8.1}",
	journal   = "Comput. Phys. Commun.",
	volume    = "178",
	year      = "2008",
	pages     = "852-867",
	eprint    = "0710.3820",
	archivePrefix = "arXiv",
	primaryClass  =  "hep-ph",
	doi       = "10.1016/j.cpc.2008.01.036",
	SLACcitation  = "%%CITATION = 0710.3820;%%"
}

@article{Sjostrand:2014zea,
      author         = "Sj{\"o}strand, Torbj{\"o}rn and Ask, Stefan and Christiansen,
                        Jesper R. and Corke, Richard and Desai, Nishita and Ilten,
                        Philip and Mrenna, Stephen and Prestel, Stefan and
                        Rasmussen, Christine O. and Skands, Peter Z.",
      title          = "{An introduction to PYTHIA 8.2}",
      journal        = "Comput. Phys. Commun.",
      volume         = "191",
      year           = "2015",
      pages          = "159",
      doi            = "10.1016/j.cpc.2015.01.024",
      eprint         = "1410.3012",
      archivePrefix  = "arXiv",
      primaryClass   = "hep-ph",
      reportNumber   = "LU-TP-14-36, MCNET-14-22, CERN-PH-TH-2014-190,
                        FERMILAB-PUB-14-316-CD, DESY-14-178, SLAC-PUB-16122,
                        --FERMILAB-PUB-14-316-CD",
      SLACcitation   = "%%CITATION = ARXIV:1410.3012;%%"
}

@Article{Lange:2001uf,
      author         = "Lange, D. J.",
      title          = "{The EvtGen particle decay simulation package}",
      booktitle      = "{Proceedings, 7th International Conference on B physics
                        at hadron machines (BEAUTY 2000)}",
      journal        = "Nucl. Instrum. Meth. A",
      volume         = "462",
      year           = "2001",
      pages          = "152",
      doi            = "10.1016/S0168-9002(01)00089-4",
      SLACcitation   = "%%CITATION = NUIMA,A462,152;%%"
}

@Article{Frixione:2008yi,
     author    = "Frixione, Stefano and Laenen, Eric and Motylinski, Patrick and White, Chris and Webber, Bryan R.",
     title     = "{Single-top hadroproduction in association with a \(W\) boson}",
     journal   = "JHEP",
     volume    = "07",
     year      = "2008",
     pages     = "029",
     eprint    = "0805.3067",
     archivePrefix = "arXiv",
     primaryClass  =  "hep-ph",
     doi       = "10.1088/1126-6708/2008/07/029"
}

@Article{Alwall:2014hca,
      author         = "Alwall, J. and Frederix, R. and Frixione, S. and Hirschi,
                        V. and Maltoni, F. and Mattelaer, O. and Shao, H. -S. and
                        Stelzer, T. and Torrielli, P. and Zaro, M.",
      title          = "{The automated computation of tree-level and
                        next-to-leading order differential cross sections, and
                        their matching to parton shower simulations}",
      journal        = "JHEP",
      volume         = "07",
      year           = "2014",
      pages          = "079",
      doi            = "10.1007/JHEP07(2014)079",
      eprint         = "1405.0301",
      archivePrefix  = "arXiv",
      primaryClass   = "hep-ph",
      reportNumber   = "CERN-PH-TH-2014-064, CP3-14-18, LPN14-066, MCNET-14-09,
                        ZU-TH-14-14",
      SLACcitation   = "%%CITATION = ARXIV:1405.0301;%%"
}

@article{Artoisenet:2012st,
      author         = "Artoisenet, Pierre and Frederix, Rikkert and Mattelaer,
                        Olivier and Rietkerk, Robbert",
      title          = "{Automatic spin-entangled decays of heavy resonances in
                        Monte Carlo simulations}",
      journal        = "JHEP",
      volume         = "03",
      pages          = "015",
      doi            = "10.1007/JHEP03(2013)015",
      year           = "2013",
      eprint         = "1212.3460",
      archivePrefix  = "arXiv",
      primaryClass   = "hep-ph",
      reportNumber   = "NIKHEF-2012-021, CERN-PH-TH-2012-329",
      SLACcitation   = "%%CITATION = ARXIV:1212.3460;%%",
}

@article{Frixione:2007zp,
      author         = "Frixione, Stefano and Laenen, Eric and Motylinski,
                        Patrick and Webber, Bryan R.",
      title          = "{Angular correlations of lepton pairs from vector boson
                        and top quark decays in Monte Carlo simulations}",
      journal        = "JHEP",
      volume         = "04",
      year           = "2007",
      pages          = "081",
      doi            = "10.1088/1126-6708/2007/04/081",
      eprint         = "hep-ph/0702198",
      archivePrefix  = "arXiv",
      OPTprimaryClass   = "HEP-PH",
      reportNumber   = "CAVENDISH-HEP-07-01, GEF-TH-09-2007, ITP-UU-07-10,
                        NIKHEF-2007-004",
      SLACcitation   = "%%CITATION = HEP-PH/0702198;%%"
}

@Article{Bahr:2008pv,
      author         = "B{\"a}hr, M. and others",
      title          = "{Herwig++ physics and manual}",
      journal        = "Eur. Phys. J. C",
      volume         = "58",
      year           = "2008",
      pages          = "639",
      doi            = "10.1140/epjc/s10052-008-0798-9",
      eprint         = "0803.0883",
      archivePrefix  = "arXiv",
      primaryClass   = "hep-ph",
      reportNumber   = "CERN-PH-TH-2008-038, CAVENDISH-HEP-08-03, KA-TP-05-2008,
                        DCPT-08-22, IPPP-08-11, CP3-08-05",
      SLACcitation   = "%%CITATION = ARXIV:0803.0883;%%"
}

@Article{Bellm:2015jjp,
      author         = "Bellm, Johannes and others",
      title          = "{Herwig 7.0/Herwig++ 3.0 release note}",
      journal        = "Eur. Phys. J. C",
      volume         = "76",
      year           = "2016",
      number         = "4",
      pages          = "196",
      doi            = "10.1140/epjc/s10052-016-4018-8",
      eprint         = "1512.01178",
      archivePrefix  = "arXiv",
      primaryClass   = "hep-ph",
      reportNumber   = "CERN-PH-TH-2015-289, MAN-HEP-2015-15, IFJPAN-IV-2015-13,
                        HERWIG-2015-01, KA-TP-18-2015, DCPT-15-142, MCNET-15-28,
                        IPPP-15-71, --HERWIG-2015-01",
      SLACcitation   = "%%CITATION = ARXIV:1512.01178;%%"
}

@Article{Nason:2004rx,
      author         = "Nason, Paolo",
      title          = "{A new method for combining NLO QCD with shower Monte Carlo algorithms}",
      journal        = "JHEP",
      volume         = "11",
      pages          = "040",
      doi            = "10.1088/1126-6708/2004/11/040",
      year           = "2004",
      eprint         = "hep-ph/0409146",
      archivePrefix  = "arXiv",
}

@Article{Frixione:2007vw,
      author         = "Frixione, Stefano and Nason, Paolo and Oleari, Carlo",
      title          = "{Matching NLO QCD computations with parton shower
                        simulations: the POWHEG method}",
      journal        = "JHEP",
      volume         = "11",
      pages          = "070",
      doi            = "10.1088/1126-6708/2007/11/070",
      year           = "2007",
      eprint         = "0709.2092",
      archivePrefix  = "arXiv",
      primaryClass   = "hep-ph",
}

@Article{Alioli:2010xd,
      author         = "Alioli, Simone and Nason, Paolo and Oleari, Carlo and Re,
                        Emanuele",
      title          = "{A general framework for implementing NLO calculations in
                        shower Monte Carlo programs: the POWHEG BOX}",
      journal        = "JHEP",
      volume         = "06",
      pages          = "043",
      doi            = "10.1007/JHEP06(2010)043",
      year           = "2010",
      eprint         = "1002.2581",
      archivePrefix  = "arXiv",
      primaryClass   = "hep-ph",
}

@article{Aliev:2010zk,
      author         = "Aliev, M. and Lacker, H. and Langenfeld, U. and Moch, S.
                        and Uwer, P. and Wiedermann, M.",
      title          = "{HATHOR -- HAdronic Top and Heavy quarks crOss section
                        calculatoR}",
      journal        = "Comput. Phys. Commun.",
      volume         = "182",
      year           = "2011",
      pages          = "1034-1046",
      doi            = "10.1016/j.cpc.2010.12.040",
      eprint         = "1007.1327",
      archivePrefix  = "arXiv",
      primaryClass   = "hep-ph",
      reportNumber   = "DESY-10-091, HU-EP-10-33, SFB-CPP-10-60",
      SLACcitation   = "%%CITATION = ARXIV:1007.1327;%%"
}

@article{Kant:2014oha,
      author         = "Kant, P. and Kind, O. M. and Kintscher, T. and Lohse, T.
                        and Martini, T. and Mölbitz, S. and Rieck, P. and Uwer, P.",
      title          = "{HatHor for single top-quark production: Updated
                        predictions and uncertainty estimates for single top-quark
                        production in hadronic collisions}",
      journal        = "Comput. Phys. Commun.",
      volume         = "191",
      year           = "2015",
      pages          = "74-89",
      doi            = "10.1016/j.cpc.2015.02.001",
      eprint         = "1406.4403",
      archivePrefix  = "arXiv",
      primaryClass   = "hep-ph",
      reportNumber   = "HU-EP-14-22",
      SLACcitation   = "%%CITATION = ARXIV:1406.4403;%%"
}

@article{Beneke:2011mq,
      author         = "Beneke, M. and Falgari, P. and Klein, S. and Schwinn, C.",
      title          = "{Hadronic top-quark pair production with NNLL threshold
                        resummation}",
      journal        = "Nucl. Phys. B",
      volume         = "855",
      year           = "2012",
      pages          = "695-741",
      doi            = "10.1016/j.nuclphysb.2011.10.021",
      eprint         = "1109.1536",
      archivePrefix  = "arXiv",
      primaryClass   = "hep-ph",
      reportNumber   = "TTK-11-38, ITP-UU-11-26, SPIN-11-19, FR-PHENO-2011-015,
                        SFB-CPP-11-49",
      SLACcitation   = "%%CITATION = ARXIV:1109.1536;%%"
}

@article{Anastasiou:2003ds,
      author         = "Anastasiou, Charalampos and Dixon, Lance and Melnikov,
                        Kirill and Petriello, Frank",
      title          = "{High-precision QCD at hadron colliders: Electroweak
                        gauge boson rapidity distributions at next-to-next-to leading order}",
      journal        = "Phys. Rev. D",
      volume         = "69",
      year           = "2004",
      pages          = "094008",
      doi            = "10.1103/PhysRevD.69.094008",
      eprint         = "hep-ph/0312266",
      archivePrefix  = "arXiv",
      reportNumber   = "SLAC-PUB-10288, UH-511-1042-03",
      SLACcitation   = "%%CITATION = HEP-PH/0312266;%%"
}

@article{Cacciari:2011hy,
      author         = "Cacciari, Matteo and Czakon, Michal and Mangano,
                        Michelangelo and Mitov, Alexander and Nason, Paolo",
      title          = "{Top-pair production at hadron colliders with
                        next-to-next-to-leading logarithmic soft-gluon
                        resummation}",
      journal        = "Phys. Lett. B",
      volume         = "710",
      pages          = "612-622",
      doi            = "10.1016/j.physletb.2012.03.013",
      year           = "2012",
      eprint         = "1111.5869",
      primaryClass   = "hep-ph",
      archivePrefix  = "arXiv",
      reportNumber   = "CERN-PH-TH-2011-277, TTK-11-54",
      SLACcitation   = "%%CITATION = ARXIV:1111.5869;%%",
}

@article{Czakon:2012zr,
      author         = "Czakon, Michal and Mitov, Alexander",
      title          = "{NNLO corrections to top-pair production at hadron
                        colliders: the all-fermionic scattering channels}",
      journal        = "JHEP",
      volume         = "12",
      pages          = "054",
      doi            = "10.1007/JHEP12(2012)054",
      year           = "2012",
      eprint         = "1207.0236",
      archivePrefix  = "arXiv",
      primaryClass   = "hep-ph",
      SLACcitation   = "%%CITATION = ARXIV:1207.0236;%%",
}

@article{Czakon:2013goa,
      author         = "Czakon, Michal and Fiedler, Paul and Mitov, Alexander",
      title          = "{Total Top-Quark Pair-Production Cross Section at
                        Hadron Colliders Through \(\mathcal{O}(\alpha_S^4)\)}",
      journal        = "Phys. Rev. Lett.",
      volume         = "110",
      pages          = "252004",
      doi            = "10.1103/PhysRevLett.110.252004",
      year           = "2013",
      eprint         = "1303.6254",
      archivePrefix  = "arXiv",
      primaryClass   = "hep-ph",
      reportNumber   = "CERN-PH-TH-2013-056, TTK-13-08",
      SLACcitation   = "%%CITATION = ARXIV:1303.6254;%%",
}

@article{Baernreuther:2012ws,
      author         = "B{\"a}rnreuther, Peter and Czakon, Michal and Mitov,
                        Alexander",
      title          = "{Percent-Level-Precision Physics at the
                  Tevatron: Next-to-Next-to-Leading Order QCD
                  Corrections to \(q \bar{q} \to t \bar{t} + X\)}",
      journal        = "Phys. Rev. Lett.",
      volume         = "109",
      pages          = "132001",
      doi            = "10.1103/PhysRevLett.109.132001",
      year           = "2012",
      eprint         = "1204.5201",
      archivePrefix  = "arXiv",
      primaryClass   = "hep-ph",
      SLACcitation   = "%%CITATION = ARXIV:1204.5201;%%",
}

@article{Frixione:2007nw,
      author         = "Frixione, Stefano and Ridolfi, Giovanni and Nason, Paolo",
      title          = "{A positive-weight next-to-leading-order Monte Carlo for
                        heavy flavour hadroproduction}",
      journal        = "JHEP",
      volume         = "09",
      pages          = "126",
      doi            = "10.1088/1126-6708/2007/09/126",
      year           = "2007",
      eprint         = "0707.3088",
      archivePrefix  = "arXiv",
      primaryClass   = "hep-ph",
      SLACcitation   = "%%CITATION = ARXIV:0707.3088;%%",
}

@Article{Hoeche:2009rj,
      author         = "H{\"o}che, Stefan and Krauss, Frank and Schumann, Steffen and Siegert, Frank",
      title          = "{QCD matrix elements and truncated showers}",
      journal        = "JHEP",
      volume         = "05",
      pages          = "053",
      doi            = "10.1088/1126-6708/2009/05/053",
      year           = "2009",
      eprint         = "0903.1219",
      archivePrefix  = "arXiv",
      primaryClass   = "hep-ph",
}

@Article{Gleisberg:2008fv,
      author         = "Gleisberg, Tanju and H{\"o}che, Stefan",
      title          = "{Comix, a new matrix element generator}",
      journal        = "JHEP",
      volume         = "12",
      pages          = "039",
      doi            = "10.1088/1126-6708/2008/12/039",
      year           = "2008",
      eprint         = "0808.3674",
      archivePrefix  = "arXiv",
      primaryClass   = "hep-ph",
}

@Article{Schumann:2007mg,
      author         = "Schumann, Steffen and Krauss, Frank",
      title          = "{A parton shower algorithm based on Catani--Seymour dipole factorisation}",
      journal        = "JHEP",
      volume         = "03",
      pages          = "038",
      doi            = "10.1088/1126-6708/2008/03/038",
      year           = "2008",
      eprint         = "0709.1027",
      archivePrefix  = "arXiv",
      primaryClass   = "hep-ph",
}

@article{Hoeche:2012yf,
      author         = "H{\"o}che, Stefan and Krauss, Frank and Sch{\"o}nherr, Marek and
                        Siegert, Frank",
      title          = "{QCD matrix elements + parton showers. The NLO case}",
      journal        = "JHEP",
      volume         = "04",
      year           = "2013",
      pages          = "027",
      doi            = "10.1007/JHEP04(2013)027",
      eprint         = "1207.5030",
      archivePrefix  = "arXiv",
      primaryClass   = "hep-ph",
      reportNumber   = "SLAC-PUB-15191, IPPP-12-52, DCPT-12-104, LPN12-081,
                        FR-PHENO-2012-017, MCNET-12-09, --FR-PHENO-2012-017",
      SLACcitation   = "%%CITATION = ARXIV:1207.5030;%%"
}

@article{Bothmann:2019yzt,
      author = "Bothmann, Enrico and others",
      title = "{Event generation with Sherpa 2.2}",
      journal = "SciPost Phys.",
      volume = "7",
      year = "2019",
      number = "3",
      pages = "034",
      doi = "10.21468/SciPostPhys.7.3.034",
      reportNumber = "FERMILAB-PUB-19-218-T, SLAC-PUB-17433, IPPP/19/42, MCNET-19-11",
      eprint = "1905.09127",
      archivePrefix = "arXiv",
      primaryClass = "hep-ph",
}

@article{Hoeche:2011fd,
      author         = "H{\"o}che, Stefan and Krauss, Frank and Sch{\"o}nherr, Marek and
                        Siegert, Frank",
      title          = "{A critical appraisal of NLO+PS matching methods}",
      journal        = "JHEP",
      volume         = "09",
      year           = "2012",
      pages          = "049",
      doi            = "10.1007/JHEP09(2012)049",
      eprint         = "1111.1220",
      archivePrefix  = "arXiv",
      primaryClass   = "hep-ph",
      reportNumber   = "SLAC-PUB-14661, IPPP-11-67, DCPT-11-134, LPN11-58,
                        FR-PHENO-2011-019, MCNET-11-24",
      SLACcitation   = "%%CITATION = ARXIV:1111.1220;%%"
}

@article{Catani:2001cc,
      author         = "Catani, S. and Krauss, F. and Webber, B. R. and Kuhn, R.",
      title          = "{QCD Matrix Elements + Parton Showers}",
      journal        = "JHEP",
      volume         = "11",
      year           = "2001",
      pages          = "063",
      doi            = "10.1088/1126-6708/2001/11/063",
      eprint         = "hep-ph/0109231",
      archivePrefix  = "arXiv",
      reportNumber   = "CERN-TH-2000-367, CAVENDISH-HEP-00-03",
      SLACcitation   = "%%CITATION = HEP-PH/0109231;%%"
}

@article{Cascioli:2011va,
      author         = "Cascioli, Fabio and Maierh{\"o}fer, Philipp and Pozzorini, Stefano",
      title          = "{Scattering Amplitudes with Open Loops}",
      journal        = "Phys. Rev. Lett.",
      volume         = "108",
      year           = "2012",
      pages          = "111601",
      doi            = "10.1103/PhysRevLett.108.111601",
      eprint         = "1111.5206",
      archivePrefix  = "arXiv",
      primaryClass   = "hep-ph",
      reportNumber   = "ZU-TH-23-11, LPN11-66",
      SLACcitation   = "%%CITATION = ARXIV:1111.5206;%%"
}

@article{Buccioni:2019sur,
    author = "Buccioni, Federico and Lang, Jean-Nicolas and Lindert, Jonas M. and Maierh{\"o}fer, Philipp and Pozzorini, Stefano and Zhang, Hantian and Zoller, Max F.",
    title = "{OpenLoops 2}",
    eprint = "1907.13071",
    archivePrefix = "arXiv",
    primaryClass = "hep-ph",
    reportNumber = "IPPP/19/62, FR-PHENO-2019-12, PSI-PR-19-15, ZU-TH 37/19",
    doi = "10.1140/epjc/s10052-019-7306-2",
    journal = "Eur. Phys. J. C",
    volume = "79",
    number = "10",
    pages = "866",
    year = "2019"
}

@article{Denner:2016kdg,
      author         = "Denner, Ansgar and Dittmaier, Stefan and Hofer, Lars",
      title          = "{\textsc{Collier}: A fortran-based complex one-loop library in
                        extended regularizations}",
      journal        = "Comput. Phys. Commun.",
      volume         = "212",
      year           = "2017",
      pages          = "220-238",
      doi            = "10.1016/j.cpc.2016.10.013",
      eprint         = "1604.06792",
      archivePrefix  = "arXiv",
      primaryClass   = "hep-ph",
      reportNumber   = "FR-PHENO-2016-003, ICCUB-16-016",
      SLACcitation   = "%%CITATION = ARXIV:1604.06792;%%"
}

@article{Czakon:2011xx,
  author         = "Czakon, Michal and Mitov, Alexander",
  title          = "{Top++: A program for the calculation of the top-pair cross-section at hadron colliders}",
  journal        = "Comput. Phys. Commun.",
  volume         = "185",
  year           = "2014",
  pages          = "2930",
  doi            = "10.1016/j.cpc.2014.06.021",
  eprint         = "1112.5675",
  archivePrefix  = "arXiv",
  primaryClass   = "hep-ph",
  reportNumber   = "CERN-PH-TH-2011-303, TTK-11-58",
}

@article{Ciccolini:2003jy,
      author         = "Ciccolini, M. L. and Dittmaier, S. and Kr{\"a}mer, M.",
      title          = "{Electroweak radiative corrections to associated \(WH\) and
                        \(ZH\) production at hadron colliders}",
      journal        = "Phys. Rev. D",
      volume         = "68",
      year           = "2003",
      pages          = "073003",
      doi            = "10.1103/PhysRevD.68.073003",
      eprint         = "hep-ph/0306234",
      archivePrefix  = "arXiv",
      %primaryClass   = "hep-ph",
      reportNumber   = "EDINBURGH-2003-05, MPI-PHT-2003-24",
      SLACcitation   = "%%CITATION = HEP-PH/0306234;%%"
}

@article{Brein:2003wg,
      author         = "Brein, Oliver and Djouadi, Abdelhak and Harlander, Robert",
      title          = "{NNLO QCD corrections to the Higgs-strahlung processes at
                        hadron colliders}",
      journal        = "Phys. Lett. B",
      volume         = "579",
      year           = "2004",
      pages          = "149-156",
      doi            = "10.1016/j.physletb.2003.10.112",
      eprint         = "hep-ph/0307206",
      archivePrefix  = "arXiv",
      XprimaryClass   = "hep-ph",
      reportNumber   = "MPP-2003-35, CERN-TH-2003-161, PM-03-16",
      SLACcitation   = "%%CITATION = HEP-PH/0307206;%%"
}

@article{Brein:2011vx,
      author         = "Brein, Oliver and Harlander, Robert V. and Wiesemann, Marius and Zirke, Tom",
      title          = "{Top-quark mediated effects in hadronic Higgs-Strahlung}",
      journal        = "Eur. Phys. J. C",
      volume         = "72",
      year           = "2012",
      pages          = "1868",
      doi            = "10.1140/epjc/s10052-012-1868-6",
      eprint         = "1111.0761",
      archivePrefix  = "arXiv",
      primaryClass   = "hep-ph",
      reportNumber   = "CERN-PH-TH-2011-268, FR-PHENO-2011-016, WUB-11-15",
      SLACcitation   = "%%CITATION = ARXIV:1111.0761;%%"
}

@article{Altenkamp:2012sx,
      author         = "Altenkamp, Lukas and Dittmaier, Stefan and Harlander,
                        Robert V. and Rzehak, Heidi and Zirke, Tom J. E.",
      title          = "{Gluon-induced Higgs-strahlung at next-to-leading order
                        QCD}",
      journal        = "JHEP",
      volume         = "02",
      year           = "2013",
      pages          = "078",
      doi            = "10.1007/JHEP02(2013)078",
      eprint         = "1211.5015",
      archivePrefix  = "arXiv",
      primaryClass   = "hep-ph",
      reportNumber   = "CERN-PH-TH-2012-312, FR-PHENO-2012-023, WUB-12-21",
      SLACcitation   = "%%CITATION = ARXIV:1211.5015;%%"
}

@article{Denner:2014cla,
      author         = "Denner, Ansgar and Dittmaier, Stefan and Kallweit, Stefan
                        and M{\"u}ck, Alexander",
      title          = "{HAWK 2.0: A Monte Carlo program for Higgs production
                        in vector-boson fusion and Higgs strahlung at hadron
                        colliders}",
      journal        = "Comput. Phys. Commun.",
      volume         = "195",
      year           = "2015",
      pages          = "161-171",
      doi            = "10.1016/j.cpc.2015.04.021",
      eprint         = "1412.5390",
      archivePrefix  = "arXiv",
      primaryClass   = "hep-ph",
      reportNumber   = "FR-PHENO-2014-013, MITP-14-101, TTK-14-36",
      SLACcitation   = "%%CITATION = ARXIV:1412.5390;%%"
}

@article{Brein:2012ne,
      author         = "Brein, Oliver and Harlander, Robert V. and Zirke, Tom J.
                        E.",
      title          = "{vh@nnlo -- Higgs Strahlung at hadron colliders}",
      journal        = "Comput. Phys. Commun.",
      volume         = "184",
      year           = "2013",
      pages          = "998-1003",
      doi            = "10.1016/j.cpc.2012.11.002",
      eprint         = "1210.5347",
      archivePrefix  = "arXiv",
      primaryClass   = "hep-ph",
      SLACcitation   = "%%CITATION = ARXIV:1210.5347;%%"
}

@article{Harlander:2014wda,
      author         = "Harlander, Robert V. and Kulesza, Anna and Theeuwes,
                        Vincent and Zirke, Tom",
      title          = "{Soft gluon resummation for gluon-induced Higgs
                        Strahlung}",
      journal        = "JHEP",
      volume         = "11",
      year           = "2014",
      pages          = "082",
      doi            = "10.1007/JHEP11(2014)082",
      eprint         = "1410.0217",
      archivePrefix  = "arXiv",
      primaryClass   = "hep-ph",
      reportNumber   = "MS-TP-14-17, WUB-14-10, LPN14-112",
      SLACcitation   = "%%CITATION = ARXIV:1410.0217;%%"
}

@article{Agostinelli:2002hh,
      author         = "Agostinelli, S. and others",
      title          = "{\textsc{Geant4} -- a simulation toolkit}",
      journal        = "Nucl. Instrum. Meth. A",
      volume         = "506",
      year           = "2003",
      pages          = "250",
      doi            = "10.1016/S0168-9002(03)01368-8",
      reportNumber   = "SLAC-PUB-9350, FERMILAB-PUB-03-339",
      SLACcitation   = "%%CITATION = NUIMA,A506,250;%%"
}

@Article{Read:2002hq,
    Author = {Read, Alexander L.},
    Journal = {J. Phys. G},
    Pages = {2693},
    doi = "10.1088/0954-3899/28/10/313",
    Title = "{Presentation of search results: the \(CL_S\) technique}",
    Volume = {28},
    Year = {2002}
}

@Article{Cowan:2010js,
   Author = {Cowan, Glen and Cranmer, Kyle and Gross, Eilam and Vitells, Ofer},
   Title = {{Asymptotic formulae for likelihood-based tests of new physics}},
   Journal = {Eur. Phys. J. C},
   Volume = {71},
   Year = {2011},
   Pages = {1554},
   doi = "10.1140/epjc/s10052-011-1554-0",
   Eprint = "1007.1727",
   Archiveprefix = {arXiv},
   Primaryclass = {physics.data-an},
   related = "Cowan:2010js-err",
   relatedstring  = "Erratum:",
}

@Booklet{RooFit,
      author         = "Verkerke, Wouter and Kirkby, David",
      title          = "{The RooFit toolkit for data modeling}",
      year           = "2003",
      eprint         = "physics/0306116",
      archivePrefix  = "arXiv",
      primaryClass   = "physics.data-an",
}


%% file: ATLAS.bib
@Article{PERF-2007-01,
    author         = "{ATLAS Collaboration}",
    title          = "{The ATLAS Experiment at the CERN Large Hadron Collider}",
    journal        = "JINST",
    volume         = "3",
    year           = "2008",
    pages          = "S08003",
    doi            = "10.1088/1748-0221/3/08/S08003",
    primaryClass   = "hep-ex",
}

@Article{SOFT-2010-01,
    author         = "{ATLAS Collaboration}",
    title          = "{The ATLAS Simulation Infrastructure}",
    journal        = "Eur. Phys. J. C",
    volume         = "70",
    year           = "2010",
    pages          = "823",
    doi            = "10.1140/epjc/s10052-010-1429-9",
    eprint         = "1005.4568",
    archivePrefix  = "arXiv",
    primaryClass   = "physics.ins-det",
}

@Article{HIGG-2013-23,
    author         = "{ATLAS Collaboration}",
    title          = "{Search for the \(b\bar{b}\) decay of the Standard Model Higgs boson in associated \((W/Z)H\) production with the ATLAS detector}",
    journal        = "JHEP",
    volume         = "01",
    year           = "2015",
    pages          = "069",
    doi            = "10.1007/JHEP01(2015)069",
    reportNumber   = "CERN-PH-EP-2014-214",
    eprint         = "1409.6212",
    archivePrefix  = "arXiv",
    primaryClass   = "hep-ex",
}

@Article{PERF-2014-03,
    author         = "{ATLAS Collaboration}",
    title          = "{Performance of pile-up mitigation techniques for jets in \(pp\) collisions at \(\sqrt{s} = 8\,\text{TeV}\) using the ATLAS detector}",
    journal        = "Eur. Phys. J. C",
    volume         = "76",
    year           = "2016",
    pages          = "581",
    doi            = "10.1140/epjc/s10052-016-4395-z",
    reportNumber   = "CERN-PH-EP-2015-206",
    eprint         = "1510.03823",
    archivePrefix  = "arXiv",
    primaryClass   = "hep-ex",
}

@Article{PERF-2014-07,
    author         = "{ATLAS Collaboration}",
    title          = "{Topological cell clustering in the ATLAS calorimeters and its performance in LHC Run~1}",
    journal        = "Eur. Phys. J. C",
    volume         = "77",
    year           = "2017",
    pages          = "490",
    doi            = "10.1140/epjc/s10052-017-5004-5",
    reportNumber   = "CERN-PH-EP-2015-304",
    eprint         = "1603.02934",
    archivePrefix  = "arXiv",
    primaryClass   = "hep-ex",
}

@Article{EXOT-2015-23,
    author         = "{ATLAS Collaboration}",
    title          = "{Search for dark matter in association with a Higgs boson decaying to \(b\)-quarks in \(pp\) collisions at \(\sqrt{s} = 13\,\text{TeV}\) with the ATLAS detector}",
    journal        = "Phys. Lett. B",
    volume         = "765",
    year           = "2017",
    pages          = "11",
    doi            = "10.1016/j.physletb.2016.11.035",
    reportNumber   = "CERN-EP-2016-181",
    eprint         = "1609.04572",
    archivePrefix  = "arXiv",
    primaryClass   = "hep-ex",
}

@Article{PERF-2015-09,
    author         = "{ATLAS Collaboration}",
    title          = "{Jet reconstruction and performance using particle flow with the ATLAS Detector}",
    journal        = "Eur. Phys. J. C",
    volume         = "77",
    year           = "2017",
    pages          = "466",
    doi            = "10.1140/epjc/s10052-017-5031-2",
    reportNumber   = "CERN-EP-2017-024",
    eprint         = "1703.10485",
    archivePrefix  = "arXiv",
    primaryClass   = "hep-ex",
}
